\documentclass[lettersize,journal]{IEEEtran}
\usepackage{amsmath,amsfonts}
\usepackage{algorithm}
\usepackage{algpseudocode}
\usepackage{dirtytalk}
\usepackage{array}
\usepackage[font=normalsize,labelfont=sf,textfont=sf]{subfig}
\usepackage{textcomp}
\usepackage{stfloats}
\usepackage{url}
\usepackage{verbatim}
\usepackage{graphicx}
\usepackage{cite}
\usepackage{multirow}
\usepackage{makecell}

\def\BibTeX{{\rm B\kern-.05em{\sc i\kern-.025em b}\kern-.08em
    T\kern-.1667em\lower.7ex\hbox{E}\kern-.125emX}}
\usepackage{balance}
\begin{document}
\title{An Exhaustive Evaluation of TTS- and VC-based Data Augmentation for ASR}
\author{Sewade Ogun \IEEEmembership{Graduate Student Member, IEEE}, 
        Vincent Colotte \IEEEmembership{Member, IEEE},
        Emmanuel Vincent \IEEEmembership{Fellow, IEEE}}
\markboth{IEEE/ACM TRANSACTIONS ON AUDIO, SPEECH, AND LANGUAGE PROCESSING, VOL. XX, NO. XX, XX~2023}%
{Author1 \textit{(et al.)}: Paper Title}

\maketitle

\begin{abstract}

Augmenting the training data of automatic speech recognition (ASR) systems with synthetic data generated by text-to-speech (TTS) or voice conversion (VC) has gained popularity in recent years. Several works have demonstrated improvements in ASR performance using this augmentation approach. However, because of the lower diversity of
synthetic speech, naively combining synthetic and real data often does not yield the best results. 
In this work, we leverage recently proposed flow-based TTS/VC models allowing greater speech diversity, and assess the respective impact of augmenting various speech attributes on the word error rate (WER) achieved by several ASR models. Pitch augmentation and VC-based speaker augmentation are found to be ineffective in our setup. Jointly augmenting all other attributes reduces the WER of a Conformer-Transducer model by 11\% relative on Common Voice and by up to 35\% relative on LibriSpeech compared to training on real data only.

\end{abstract}



\section{Introduction}

Training data augmentation is a popular approach for improving the performance of automatic speech recognition (ASR) systems. It makes it possible to learn features that are more invariant to perturbations and to synthesise testing conditions that are not covered in the real training data. Prominent augmentation methods include time and frequency masking \cite{park19e_interspeech}, speed change \cite{ko2015audio}, and noise addition \cite{kim17_interspeech}. In recent years, text-to-speech (TTS) and voice conversion (VC) based synthetic data augmentation have gained popularity due to their increasing quality and naturalness, which is becoming indistinguishable from real speech \cite{kim2021conditional, tan2022naturalspeech}. Synthetic data augmentation has been explored for various ASR applications such as keyword detection \cite{hu2022synt++}, ASR for code-switched \cite{abs-2010-05549} or low-resourced languages \cite{zevallos22_interspeech, baas2022lowresource,robinson22_interspeech}, children's ASR \cite{shahnawazuddin2020voice}, or improved numeric sequence recognition \cite{peyser19_interspeech}. While these works have proven that synthetic data can improve performance in different domains, naively generated synthetic data does not always yield the best results \cite{hu2022synt++, sharma2020improving} because of the distributional gap between real and synthetic speech \cite{minixhofer23_interspeech}.

One key to closing this gap is to generate synthetic data with similar diversity to real speech \cite{hu2022synt++}. The characteristics of an ASR dataset that make it diverse are manifold. Phonetic coverage is important 
to correctly recognise uncommon words \cite{wu2007data}. Also, diverse speakers ensure that the ASR model is invariant to individual voice characteristics. Similarly, attributes such as speaking style, emotion, or stress result in variations in prosody \cite{kim2021conditional}. 
Finally, various types of noises and reverberation ensure robustness to the recording conditions \cite{minixhofer23_interspeech}. Accounting for all these attributes is required for synthetic data to match or even fill in holes in the real data distribution.

To the best of our knowledge, only the works in \cite{rossenbach2023relevance} and \cite{minixhofer23_interspeech} have recently examined the impact of acoustic diversity of synthesised speech on the final ASR results, and no recent work has explored the impact of linguistic diversity. 
Therefore, in this work, we perform an experimental study to measure the performance gains that can be obtained from synthetic data augmentation by varying the speech attributes in the dataset. In particular, we independently and jointly evaluate the impact of the phonetic content, phoneme duration, pitch, number of speakers, and the environmental characteristics of the synthetic data for different data sizes on the ASR performance. 

Recent multi-speaker TTS systems can vary the acoustic characteristics of generated speech when conditioned on a specific prosodic attribute \cite{ren2020fastspeech, 9413889}. 
In particular, generative model-based TTS models learn the distribution of the acoustic and linguistic characteristics of speech from training data, which gives them the intrinsic ability to generate diverse utterances while being able to control the prosodic attributes of the speech implicitly \cite{kim2020glow} or explicitly \cite{ogun23_interspeech}. Therefore, for data generation, we leverage flow-based TTS models \cite{kim2020glow, ogun23_interspeech} which are capable of generating diverse and natural utterances.

The rest of the paper is organised as follows. Section~\ref{related_work} highlights the works related to this experimental study. Section~\ref{experiment_design} describes the TTS/VC and ASR models used in this work as well as our experimental design. We explain the experimental setup and results in Section~\ref{methodology} and conclude in Section~\ref{conclusion}.

\section{Related Work}
\label{related_work}

In this section, we review the background of this paper in the following fields: TTS, VC, and ASR data augmentation.

\subsection{Related TTS Research}

TTS involves generating speech from text (or other linguistic features). Early TTS systems include concatenative corpus-based systems \cite{hunt1996unit} and Hidden Markov model (HMM) based statistical parametric speech synthesis systems  \cite{tokuda2002hmm} which produce less natural synthetic speech than current deep-learning-based TTS systems \cite{tan2021survey}.

Deep-learning-based methods have mostly eliminated human knowledge from the TTS system's design. 
For example, the older text analysis component of TTS models has been replaced by a module that learns contextual text representation through self-attention. 
Deep-learning-based TTS models can be broadly classified into multi-stage TTS systems and end-to-end TTS systems. Multi-stage TTS systems take tokenised text such as phonemes or characters as input and produce an intermediate representation such as a Mel-spectrogram \cite{ren2020fastspeech}. The intermediate representation is then converted to speech using a vocoder. The vocoder either approximates the phase of the spectrogram as in iterative phase retrieval methods such as the Griffin-Lim algorithm \cite{perraudin2013fast} or directly generates the audio given the spectrogram using deep-learning-based methods \cite{vandenoord16_ssw, kong2020hifi}. Deep-learning-based vocoders synthesise more natural and intelligible utterances than classical phase-retrieval methods and are therefore the common choice of vocoders in the last decade. 
End-to-end TTS systems merge the entire TTS pipeline by generating speech directly from the text input \cite{kim2021conditional}. 

A TTS system can also be either single-speaker or multi-speaker. A single-speaker TTS system generates speech in the voice of the speaker it has been trained with. It is desirable to be able to synthesise natural speech for multiple speakers using a single system. Therefore,
several research focuses on improving and controlling speaker attributes like emotion, speaking style, etc., in multi-speaker TTS systems. 
Multi-speaker TTS systems typically add extra speaker-conditioning to a TTS system to learn the speaker characteristics during training and to generate speech in the voice of the speaker during inference. The speaker-conditioning can be a speaker embedding that is trained alongside the TTS system \cite{ren2020fastspeech} or an external speaker embedding extracted from a speaker recognition system \cite{casanova2021sc}. With an external speaker embedding, it is possible to synthesise speech in the voice of a speaker unseen during training, although this is a more challenging task. 

Additionally, TTS systems can either make use of discriminative or generative modelling approaches for acoustic modelling. While non-generative TTS models assume a one-to-one mapping between the input text and the speech \cite{ren2020fastspeech}, which does not match the intrinsic diversity of real-world speech, generative deep-learning models learn the distribution of the training data during training and sample from their latent space to generate diverse data \cite{Zhang2023}. Generative TTS models include TTS models based on Normalizing Flows \cite{kim2020glow, kim2021conditional, shih2021rad}, Diffusion models \cite{popov2021grad}, and Variational Autoencoders (VAE) \cite{9053436}. 
Flow-based generative models, in particular, can fit complex data distributions and generate diverse utterances with high quality. They have also been shown to be on par with recent diffusion probabilistic models for acoustic modelling \cite{zhang23o_interspeech}.

Several attributes in speech are important for improving the perceived quality and diversity of speech. Speech attributes such as pitch, energy, phoneme duration, and rhythm can be modelled in the TTS system implicitly or explicitly. Explicitly learning the attributes is particularly useful for independently changing the characteristics of the generated speech. The works of \cite{ogun23_interspeech} and \cite{lee2022varianceflow} have recently explored using stochastic flow-based models to learn these variabilities. Using flow-based models for predicting the pitch and phoneme duration also increases the diversity of the utterances generated by the TTS models \cite{ogun23_interspeech, shih2021rad}, since in multi-speaker TTS synthesis, speakers can have different speaking styles, pronunciations, speaking rates, and intonations, which can be directly learned by the predictors. We have used flow-based TTS models with and without external predictors in our experiments to be able to generate utterances with controllable attributes. Our predictors are particularly generative flow-based predictors which provide the model with diverse speech attributes during generation.

\subsection{Related VC research}

VC is the process of converting a reference speaker's voice into a target speaker's voice while preserving the linguistic information \cite{sisman2020overview}. Depending on the system, it may preserve the prosody of speech while changing the timbre of the speaker \cite{kim2020glow, sisman2020overview}. The transformations between the source speaker and target speaker can be learned from parallel data or non-parallel
data depending on the voice conversion approach.

Similar to TTS research, early statistical approaches to VC include Gaussian mixture models (GMMs) \cite{toda2007voice}, negative matrix factorisation (NMF) \cite{takashima2012exemplar}, HMMs \cite{ye2006quality}, partial least square regression \cite{helander2010voice}, and dynamic kernel partial least squares regression (DKPLS) \cite{helander2011voice}. For the methods, the speech quality
and similarity to the target speaker were not high. Deep-learning-based methods have
advanced the state-of-the-art for voice conversion methods and have greatly improved the naturalness of generated utterances for the target speaker.

Current deep-learning-based methods compress the representation of speech to extract only
the linguistic content (and possibly the prosodic content), providing the
decoder of the VC model missing content through a speaker representation. Therefore, the architecture typically takes the form of an autoencoder and can convert any speaker's voice into any other speaker's voice \cite{qian2019autovc} given the speaker's representation. 

Popular VC architectures include VAEs \cite{qian2019autovc}, Generative Adversarial Networks (GANs) \cite{wang2020one, li21e_interspeech}, Normalizing Flows \cite{merritt2022text, li2023freevc}, and vector quantised VC models \cite{wang2021vqmivc}. Modern VC systems do not require parallel data between the source and target speaker, however, a multi-speaker dataset is usually still required for training the VC model.

There is a current wave of prompt-guided large-scale speech generation models capable of audio generation, speech completion, and style conversion e.g., \cite{borsos2023audiolm, le2306voicebox}. These models typically use self-supervised audio representations, and rely on large-scale datasets in high-resource languages like English for training the models. In our experiments, we focus on using flow-based TTS models for voice conversion as these systems do not require a large amount of data and training~h to learn the distribution of data. They can also be trained jointly with the TTS system, and have been shown to generate natural voices given the target speaker representation \cite{mosinski2023ae, kim2020glow}.

\subsection{ASR Data Augmentation using synthetic data}

ASR training data augmentation using synthetic data is mostly done by combining real speech data and synthetic speech data during model training \cite{hu2022synt++}. The synthetic data can either be TTS generated or voice converted. Synthetic data are samples generated from a finite distribution of real data that was learned during TTS/VC training. Hence, a distributional mismatch between real and synthetic data \cite{hu2022synt++, sharma2020improving}. The mismatch is usually accounted for in several ways during the ASR training phase.

The first category of data augmentation modifies the synthetic data during ASR training. Examples include noise addition \cite{minixhofer23_interspeech}, SpecAugment \cite{park19e_interspeech}, and mixup augmentation \cite{sharma2020improving}. Adding noise to the synthetic speech simulates similar environmental conditions to the real speech during training. Mixup augmentation mixes the real and synthetic speech using a mixing ratio to create a more robust feature for the input data. In the case of mixup augmentation, the model is required to be able to label both the real and synthetic mixture at its output. Similarly, a rejection sampling strategy was used by \cite{hu2022synt++} to reject samples outside the distribution of typical real speech. Modifying the data directly increases the model's robustness to noise and environmental conditions, however it is data-centric.

Another technique is to modify the ASR model's layers to account for the differences between the real and the synthetic datasets. For example, a separate batch normalisation statistic can be learned for the real and synthetic dataset \cite{hu2022synt++}. In this case, the batch normalisation statistics of the real dataset are used during inference while those of the synthetic dataset are discarded. The affine parameters of the statistics are still shared between the real and synthetic speech, which helps to reduce model overfitting.  Changing the model layers increases robustness, however it still requires some knowledge of the model architecture and several modifications to determine which layers are most important for the task.

Lastly, modifying the training objective can be used to account for both the real and synthetic domains. For instance, a consistency loss was applied by \cite{wang2020improving} to ensure similar grapheme predictions for real and synthetic speech when the synthetic data was generated using the real data texts. Also, speech chain \cite{tjandra2020machine}, which involves performing both ASR training and TTS training in a closed loop, using one part of the process to improve the other part, is another method that ensures consistency between the real and generated synthetic speech. It allows the ASR model and TTS model to learn from each other in a feedback loop. This method is quite experimental and requires the weight of the loss functions to be determined experimentally to get the best performance.

To focus on the data diversity and its impact on ASR performance in this work, we apply methods that modify the synthetic data directly to increase data diversity.

\subsection{Influence of synthetic data diversity}

Only two works that we know of in literature have explored the influence of diversity in the generated speech. The authors in \cite{rossenbach2023relevance} explored the impact of phone duration diversity of the synthetic audio, which they proposed to improve by changing the predicted duration using a heuristic random walk algorithm. A recent work \cite{minixhofer23_interspeech} also explored the diversity of speech utterances in several dimensions, however, their work focused on measuring the distribution gap between real and synthetic utterances using the relative word error rate (WER) of ASR models trained on real and synthetic speech corpora.

\section{Experimental Design}
\label{experiment_design}

This section describes the datasets, the TTS/VC models, and the ASR models used in our experiments. The procedure for systematically generating the synthetic data is also discussed in detail.

\subsection{Datasets}

ASR datasets like Common Voice contain utterances with varying ambient noise from several speakers. This is a good attribute for general ASR, however noise can degrade TTS. For TTS/VC training, it is common to select a subset of the utterances using a quality criterion \cite{ogun2023can, zen2019libritts}. ASR datasets are the kind of data that is commonly available in most practical scenarios and languages, therefore we demonstrate how a noisy, multi-speaker ASR dataset can be useful for training TTS/VC systems for ASR data augmentation.

Here, we base our study on the Common Voice (CV) dataset \cite{ardila2020common}. 
It is a crowdsourced, Creative Commons Zero licensed, read speech dataset. It contains recordings from volunteers who read a text transcript sourced from public domain text. Recordings are split into a validated and invalidated set. The validated set contains clips that have been marked with more up-votes than down-votes. We use the validated English subset of CV (version 7.0) in this work while excluding the predefined development and test sets from training both the ASR and TTS/VC systems. They were only used for evaluating the performance of our trained ASR models.

Our TTS/VC models are trained on a subset of the dataset which has been curated for TTS purposes using the approach described in \cite{ogun2023can}. This involves selecting good-quality audio recordings from the CV dataset by filtering speakers with high automatically estimated mean opinion scores. The curated TTS dataset contains 4,469 speakers with a total duration of 230.75~h. 

For our experiments, we consider the situation where the ASR dataset available in the language or domain is very small. This is typically the case in children’s ASR applications or low-resourced languages. Hence, the size of the real ASR training data was limited to
50~h or 100~h which enables us to evaluate the performance of the methods in the
low data regime.
The ASR data also contains only speakers that were used to train the TTS system to efficiently evaluate the contribution of speaker attributes.

Alongside the test set of CV, we also perform evaluations on the test-clean (LS-C) and test-other (LS-O) sets of the LibriSpeech dataset \cite{panayotov2015librispeech} to compare performance improvements across multiple datasets. In this scenario in particular, the LibriSpeech dataset is of higher quality than CV, which helps us to report results on both noisy and clean data scenarios.



\subsubsection{Data preprocessing for TTS training}
 
The recordings in our TTS dataset were preprocessed by resampling from 32~kHz or 48~kHz to 16~kHz sampling rate, and removing beginning and end silences using pydub\footnote{\url{https://github.com/jiaaro/pydub}} with a threshold of -50~dBFS.

We also processed the paired text by performing text normalisation and text tokenisation. In text normalisation, numbers and abbreviations are converted to their written forms. We keep the case of letters and common punctuation like \say{,.'!}. Text normalisation was performed using the NeMo text-processing toolkit \cite{kuchaiev2019nemo}. For text tokenisation, during training the texts were converted into phonemes (with stress markers excluded) and characters in a mixed approach. With a probability of $0.2$, a word in a sentence is tokenised into characters or replaced with its phonetic representation. This improves the robustness of the TTS model to words without phoneme mapping or to out-of-vocabulary words. 

\subsubsection{Data preprocessing for ASR training}

Our ASR real training dataset contains only 2,457 speakers which are the speakers that intersect with the TTS data speakers in the CV training set. We convert all the recordings to 16~kHz wav files from mp3 files. The remaining speakers in the training set were used as candidate ASR unseen speakers in our experiments.

In addition, we generated synthetic utterances using three TTS/VC models (described in Section \ref{tts_models}). In the experiments covering individual attributes, we either generated 50~h or 100~h of synthetic data using the TTS/VC models. When we combined several speech attributes in the final experiments, we increased the size of the data until we did not see any improvements. Voice conversion was also performed on
the real training data for data augmentation. Depending on the experiment, the speakers used as output targets by TTS/VC can either be speakers in the real ASR data or new speakers.
Table~\ref{tab:table_2} summarises the data duration and the number of speakers used in this experimental study.

\begin{table}
\caption{Data duration and number of speakers used for training the ASR and TTS/VC systems. We continuously increase the number of speakers by a factor of 2 for speaker diversity experiments described in Section~\ref{speaker-diversity}.}\label{tab:table_2}
\centering
\begin{tabular}{| l | l | l | l |}
\hline
Models & Real/synthetic & Duration (h)& Speakers \\
\hline
TTS & real & 230.75 & 4,469 \\
\hline
ASR & real & 50 & 2,457 \\
\hline
ASR & real & 100 & 2,457 \\
\hline
ASR & synthetic & 50 & 2,457 \\
\hline
ASR & synthetic & 100 & 2,457 (x2, x4, x8) \\
\hline

\end{tabular}
\end{table}

\subsection{Speech synthesis models}
\label{tts_models}
\begin{figure}[!htbp]
\centering
\includegraphics[width=\linewidth]{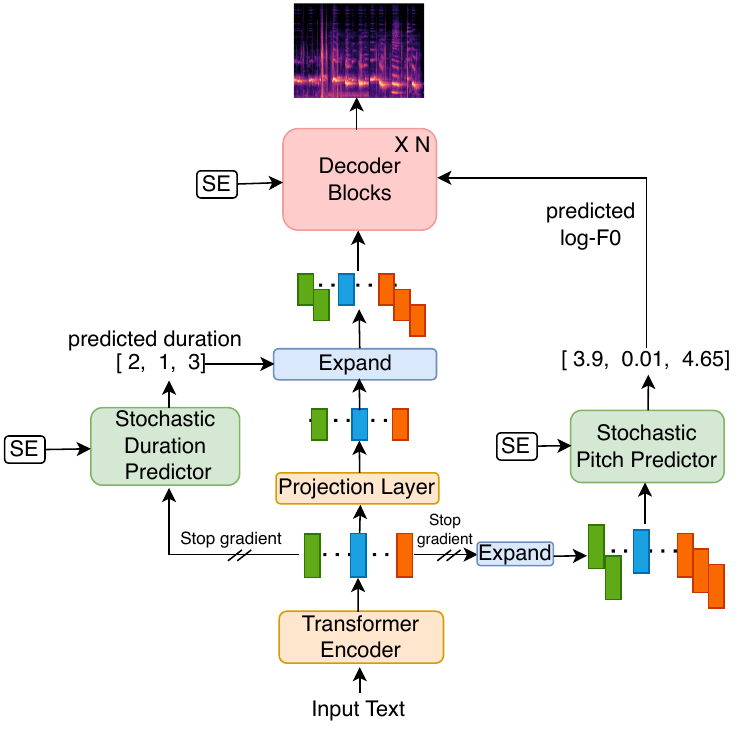}
\caption{GlowTTS-STDP architecture during inference \cite{ogun23_interspeech}.}
\label{fig_glowtts_stdp}
\end{figure}

To create synthetic data with different speech characteristics, we trained three different types of TTS/VC models: the classical speaker-conditioned Glow-TTS \cite{casanova2021sc} and two other recently proposed flow-based models, GlowTTS-STD and GlowTTS-STDP \cite{ogun23_interspeech}. The speaker-conditioned Glow-TTS model is a multi-speaker, non-autoregressive, flow-based generative TTS/VC model. The model uses a Transformer encoder and a flow-based decoder, along with a deterministic phoneme duration prediction network.
The GlowTTS-STD model is a Glow-TTS model enhanced with an improved flow-based duration predictor. The stochastic duration predictor enables the model to have a more human-like rhythm than the monotonic rhythm of the classical Glow-TTS model. The stochastic duration predictor is composed of four normalizing flow stages that learn the discrete duration estimated using the Monontonic Alignment Search (MAS) algorithm proposed in Glow-TTS. Similar to GlowTTS-STD, the GlowTTS-STDP model shown in Figure~\ref{fig_glowtts_stdp} incorporates a stochastic duration predictor and a stochastic pitch predictor into the TTS system. The pitch prediction model is also composed of four normalizing flow stages that learn the distribution of the log-F0 values for each phoneme conditioned on the speaker. The GlowTTS-STDP decoder is additionally conditioned on the pitch information of the utterance, in the form of log-F0 values. The log-F0 values extracted from the ground-truth Mel-spectrograms are used as pitch targets for training the flow-based pitch predictor and for conditioning the decoder.

To generate utterances during inference, the contextual tokenised text embeddings from the encoder are projected using a projection layer, which is then expanded using the phoneme duration predicted by the duration prediction. Alongside, the contextual tokenised text embeddings are used by the stochastic pitch predictor to generate log-F0 values (in the case of GlowTTS-STDP). The stochastic duration and stochastic pitch predictors improve the diversity of generated utterances as reported in previous listening tests \cite{ogun23_interspeech}. In our experiments, we evaluate if this improvement in the diversity of utterances also translates to improving the performance of ASR models.

\begin{figure}[!ht]
\centering
\captionsetup[subfloat]{labelfont=scriptsize,textfont=scriptsize}
\subfloat[Mel-spectrogram inversion]{\includegraphics[width=1.7in,trim=0 0 0 0,clip]{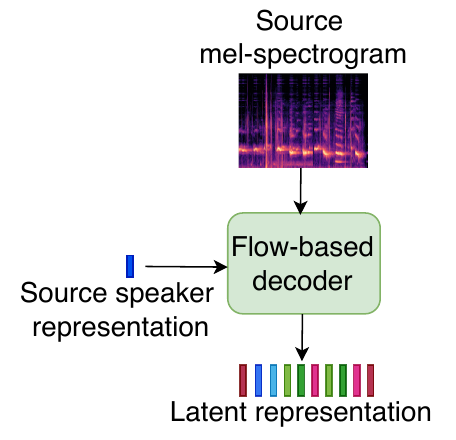}\label{fig:glowtts-vc-inversion}}
\subfloat[Voice conversion]{\includegraphics[width=1.7in,trim=0 0 0 0cm,clip]{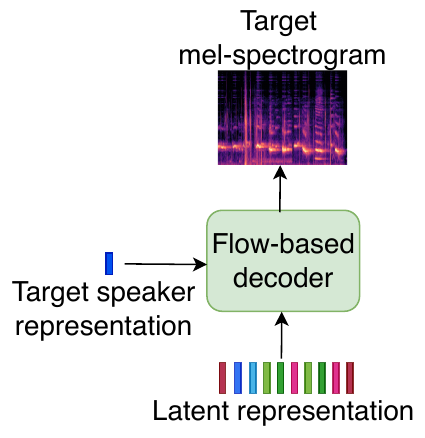}\label{fig:glowtts-vc-conversion}}
\caption[Flow-based VC showing the Mel-spectrogram inversion and voice conversion processes.]{Flow-based VC showing the Mel-spectrogram inversion and voice conversion (VC) processes. \label{fig:glowtts-vc}}
\end{figure}

The VC process is rather simple for the Glow-TTS and GlowTTS-STD models. As shown in Figure~\ref{fig:glowtts-vc}, the Mel-spectrogram of the source speaker is converted into latent representations using the flow-based decoder, then converted back into a Mel-spectrogram while conditioning the latent representations on the new speaker's representation. New phoneme durations may be predicted for the latent distribution using the model's duration predictor. In addition, the GlowTTS-STDP model also conditions the representation on the log-F0 values predicted for the new speaker when inverting the representation to generate a Mel-spectrogram.

A HiFi-GAN V1 \cite{kong2020hifi} vocoder was used to convert the generated Mel-spectrograms to speech at a 16~kHz sampling rate, which is suitable for ASR training. The vocoder has been trained on the LibriTTS dataset.

Speaker representation is provided through speaker embedding extracted from a speaker extraction model, Resemblyzer\cite{wan2018generalized}.\footnote{\url{https://github.com/resemble-ai/Resemblyzer}} For each speaker, we extract embeddings for multiple utterances and average the speaker's utterance embeddings to compute a speaker representation for inference.

\subsection{ASR model}

ASR models take speech features as inputs and output characters or words. We used two ASR models in this study, a state-of-the-art Conformer-Transducer (CTD) \cite{gulati20_interspeech} and a wav2vec2 ASR model \cite{baevski2020wav2vec}. The CTD model was used for evaluating the speech attributes that are important for improving the performance of ASR while the wav2vec2 model was finetuned on the final datasets generated after combining all the speech attributes to create a diverse synthetic dataset. The CTD model is composed of a convolution-augmented Transformer (Conformer) encoder, a Long Short Term Memory (LSTM) decoder, and a fully connected layer as a joint network. The Conformer encoder sandwiches convolutional layers between self-attention blocks. The convolutional layers better capture local information while the self-attention layers help to capture the global information from the hidden features. The encoder consists of 16 layers of Conformer blocks with 4 attention heads and a hidden dimension of 256. A 1-layer LSTM is used as the main decoder with the recurrent neural network transducer (RNNT) loss \cite{graves2012sequence}. Following standard practice, we use an auxiliary Connectionist Temporal Classification (CTC) loss \cite{graves2006connectionist} obtained through an extra convolution decoder on top of the encoder.
Hence, we combine two losses to derive the ASR training loss $\mathcal{L}_\text{ASR}$ as
\begin{equation}
\label{conformer_loss}
\mathcal{L}_\text{ASR} = (1 - \alpha)\,\mathcal{L}_\text{RNNT} + \alpha\,\mathcal{L}_\text{CTC},
\end{equation}
where $\mathcal{L}_\text{RNNT}$ is the main loss from the transducer and $\mathcal{L}_\text{CTC}$ is the auxiliary loss from the CTC branch. The model contains 31.2~M trainable parameters. The wav2vec2 model is an acoustic model pre-trained with a self-supervised objective. We selected the model trained on the 100{,}000~h unlabeled subset of Voxpopuli dataset \cite{wang-etal-2021-voxpopuli}.\footnote{\url{https://huggingface.co/facebook/wav2vec2-large-100k-voxpopuli}} The model serves as a feature extraction module with a randomly initialised decoder finetuned using a CTC objective on top of the encoder. We have used a pre-trained model for our final experiments since pre-trained models have been shown to perform better than models trained from scratch for low-resourced datasets. In our experiments, we freeze the feature extractor since this reduces the training time. The wav2vec2 model contains 317~M parameters.

\subsection{Training and inference hyper-parameters}

The CTD ASR model and TTS models\footnote{Code for training TTS models can be found at~\url{https://github.com/ogunlao/glowtts_stdp}} were trained using the NeMo toolkit \cite{kuchaiev2019nemo} while the wav2vec2 model was trained using Hugging Face Transformers library \cite{wolf-etal-2020-transformers} with Deepspeed \cite{deepspeed} for multi-gpu training on 4 GPUs. The CTD ASR model uses 80-dimensional log-Mel features as inputs and outputs 26 English alphabetical characters, space, and apostrophe.
  
Each CTD ASR model was trained for 500 epochs with early stopping on the validation set after 50 epochs while the wav2vec2 model was trained for 100 epochs. The AdamW optimizer and the NoamAnnealing scheduler with a linear warm-up of 6,000 iterations were used for gradient optimisation. SpecAugment \cite{park19e_interspeech} was also applied on the input features. We report the word error rate (WER) and character error rate (CER) on the CV test set for each ASR model and compare these results to the CER and WER on LS-C and LS-O sets of the LibriSpeech dataset. 
The $\alpha$ value for combining the CTD losses was set to $0.3$. No external language model was used for decoding.

To compare the performance of the ASR models, we performed statistical tests using the Matched Pairs Sentence-Segment Word Error (MAPSSWE) Test \cite{gillick1989some} implemented in NIST's scoring toolkit. It is a standard parametric test that looks at the number of errors in segments of utterances of varying size that are specific to the output of the two ASR systems being compared. Bold values in the tables indicate the best system (for each combination of real and synthetic data durations) and those systems that are statistically equivalent to it at a 95\% confidence level.

For TTS inference, the noise temperature of the prior distribution was set to 0.667. The noise temperature for the pitch predictor and duration predictor of GlowTTS-STD and GlowTTS-STDP was set to $0.8$. The noise temperatures control the weight of the additive Gaussian noise added to the latent representations in each flow-based model.

\section{Experimental methodology and results}
\label{methodology}

The experiments were divided into subsections covering each of the factors we plan to evaluate and improve upon. The first subsection focuses on the linguistic content of speech, while the following subsections evaluate the impact of the speaker or prosodic characteristics, i.e., speakers, pitch, and phoneme duration. Then, we evaluate the impact of adding noise to the synthetic utterances. In the last subsection, we combine all the factors to create a diverse synthetic dataset for ASR training data augmentation. Each subsection describes the methodology used to generate the synthetic dataset and the corresponding ASR results.

For the ASR training data, we use the notation \textit{A}r\_\textit{B}tts or \textit{A}r\_\textit{B}vc to denote that \textit{A}~h of real speech and \textit{B}~h of TTS/VC data were used in the specific experiment, e.g., 50r\_50tts means that 50~h of real ASR data and 50~h of TTS data were combined for ASR training. 

\subsection{Increasing phonetic diversity}
 
For languages or domains where a text corpus is more readily available than real speech data with corresponding texts, it may be useful to generate synthetic data to increase the phonetic content of the real ASR training data using the available texts. In our experiments, we applied an iterative greedy method to select sentences that we used to generate TTS data. The method is described below.

\subsubsection{Methodology}
Let $X = \{x_1, x_2, x_3, ..., x_n\}$ be a large set containing $n$ sentences that covers the natural distribution of text in a specific language or domain. If a subset $Y \subset X$ of sentences (and the corresponding real utterances) is available as real training data, then the remaining texts in $Z = X \setminus Y$ are candidate sentences for generating synthetic utterances. Our goal is to select a subset of sentences $K \subset Z$ given a constraint on the size of the TTS data (e.g., a duration of $T_\text{s}$~h) that minimises the Kullback-Leibler (KL) divergence between the distribution of the combined real and synthetic data $Y \cup K$ and a desired distribution. 
In practice, we consider the distribution of pairs of adjacent phonemes, called di-phonemes, which account for the left or right context of every phoneme. Although it is also possible to use the distribution of tri-phonemes or word units, when $Z$ is very large, the choice of di-phonemes increases the ratio of common to uncommon units. 

Concretely, given a dataset $D$ with all its sentences split into di-phonemes, for any di-phoneme $d_i \in D$, we compute its probability $P_D(d_i)$ in dataset $D$.
\begin{equation}
\label{diph_prob}
P_D(d_i) = \frac{\text{count of di-phoneme}~d_i~\text{in }D}{\text{total no. of di-phonemes in }D}.
\end{equation}
For discrete distributions and given an arbitrary set $X$, the KL divergence is defined as: 
\begin{equation}
\text{KL}(P||Q) = \sum_{c \in \text{C}} P(c) \left(\log P(c) - \log Q(c) \right),
\end{equation}
where $P(c)$ is the empirical distribution and $Q(c)$ is the model distribution. Usually, $P(c)$ is a fixed dataset and $Q(c)$ is a model whose parameters we want to adapt. In our case, it is the opposite: $Q(c)$ is a fixed desired distribution and we adapt the dataset distribution $P(c)$ to best fit it.
Therefore, the KL divergence between the distribution of di-phonemes in $Y \cup K$ and the desired distribution $Q$ becomes
\begin{equation}
\label{kl_div}
\text{KL}(P_{Y \cup K}||Q) = \sum_{d_i \in \text{diph}} P_{Y \cup K}(d_i) (\log P_{Y \cup K}(d_i) -\log Q(d_i) ).
\end{equation}
In the following, we consider two choices for the desired distribution $Q(d_i)$: the \emph{natural} distribution $P_X(d_i)$ of di-phonemes in $X$ and \emph{uniform} distribution, that is, the distribution with maximum entropy recommended by \cite{wu2007data}.

We iteratively minimise the divergence by selecting utterances one at a time using Algorithm~\ref{alg:alg_scp}. The set of selected sentences $K$ is used to synthesise speech, and the ASR system is then trained on the dataset corresponding to $Y \cup K$. 

\begin{algorithm}[H]
\caption{Algorithm for sentence selection.}\label{alg:alg_scp}
\begin{algorithmic}[1]
\Require $T_\text{s}$, $Z$, $Q$ 
\Procedure{SelectSentence}{$T_\text{s}, X, Y$}
    \State Initialise $K = \emptyset,\ T_\text{c} = 0$ \Comment{$T_\text{c}$ is the total duration of current data selected}
    \While{$T_\text{c} < T_\text{s}$}
        \For{each sentence $z$ in $Z$}
        \State \hspace{0.5cm} Compute~$\text{KL}_z := \text{KL} \left(P_{Y \cup K \cup z} || Q \right)$
        \EndFor
    \State $z_\text{s}$ := $z$~with minimum $\text{KL}_z$ 
    \State $Z \gets Z \setminus \{z_\text{s}\}$,\ $K \gets K \cup \{z_\text{s}\}$
    \State $T_\text{c}:= T_\text{c} + \text{duration of clip corresponding to}~z_\text{s}$
    \EndWhile
    \State return $K$
\EndProcedure
\end{algorithmic}
\end{algorithm}

The set of sentences in the CV training set was used as $X$ and di-phonemes were computed by first extracting the sequence of phonemes corresponding to the text using a grapheme-to-phoneme converter,\footnote{\url{https://github.com/Kyubyong/g2p}} and then consider pairs of adjacent phonemes. 
In addition, we use the duration of real utterances corresponding to the text during iteration as that would not require us to generate TTS utterances when running the algorithm.
Lastly, we optimise the algorithm to run on a 40-core CPU which takes about two days to select text equivalent to 50~h of speech.

\subsubsection{Results}

In Table~\ref{tab:tablea}, we provide a baseline for CTD ASR models trained on 50~h and 100~h of real data. Additionally, we provide a baseline (50r\_50ss) where the 50 hour TTS dataset was generated using sentences from the real data. The CER and WER of the baselines are higher than typical values reported for these data sizes due to the data selection method which restricts the selection of speakers to only speakers that intersect with the TTS training data. As such the ASR training dataset is "high-quality" but then makes it different from the significantly noisier CV test set. This is reflected in the relatively lower CER and WER of the ASR models on LibriSpeech test sets.

We compare the proposed sentence selection method using the natural distribution of $X$ or the uniform distribution as a target to \emph{random} selection of sentences from set $Z$. For 50~h of TTS data, the algorithm selects $24{,}879$, $28{,}664$, and $24{,}879$ sentences for natural, uniform, and random selection methods respectively.
Table~\ref{tab:tablea} shows the results for CTD ASR models trained on utterances generated from 50~h or 100~h of selected sentences combined with 50~h or 100~h of real speech. 
The classical speaker-conditioned Glow-TTS model was used to generate the synthetic utterances for the $2{,}457$ speakers in the real data. We cover all the speakers in the set during synthesis by continuously iterating through the set of $2{,}457$ ASR speakers until the size of the TTS dataset has been reached.
\begin{table}[!ht]
\caption{\%~CER and \%~WER of CTD ASR models trained on real speech and TTS speech generated from sentences selected randomly or using the uniform or natural distribution criterion. Bold numbers denote the best selection method and the systems statistically equivalent to it for each amount of data.}\label{tab:tablea}
\centering

\begin{tabular}{ l  l | l  l | l  l }
\hline
\multirow{2}*{\makecell{Selection \\method}} & \multirow{2}*{Data} & \multicolumn{2}{c|}{CV} & LS-C & LS-O \\
\cline{3-6}
 &  & CER & WER & WER & WER \\
\hline
- & 50r & 25.01 & 46.30 & 21.18 & 37.74 \\
Real sent. & 50r\_50ss & 23.19 & 43.44 & 20.32 & 35.81 \\
\hline \hline
random & \multirow{3}{*}{50r\_50tts} & 23.28 & 42.79 & 18.70 & 35.19 \\
uniform &  & 22.56 & 42.13 & 19.35 & 35.23 \\
natural &  & \textbf{22.24} & \textbf{41.52} & \textbf{17.90} & \textbf{33.66} \\
\hline
random & \multirow{3}{*}{50r\_100tts} & 22.23 & 40.98 & \textbf{16.52} & \textbf{31.92} \\
uniform &  & 22.83 & 41.92 & 17.71 & 33.65 \\
natural &  & \textbf{21.91} & \textbf{39.89} & \textbf{16.65} & \textbf{32.12} \\
\hline
\hline
- & 100r & 22.10 & 40.81 & 16.14 & 30.98 \\
\hline \hline
random & \multirow{3}{*}{100r\_50tts} & 20.50 & 38.25 & \textbf{13.78} & \textbf{27.33} \\
uniform &  & 20.11 & 37.48 & \textbf{13.91} & \textbf{27.48} \\
natural &  & \textbf{20.00} & \textbf{37.11} & \textbf{13.94} & \textbf{27.44} \\
\hline
random & \multirow{3}{*}{100r\_100tts} & 20.21 & 37.09 & \textbf{13.27} & \textbf{27.50} \\
uniform &  & \textbf{20.00} & 36.92 & \textbf{13.38} & \textbf{27.36} \\
natural &  & \textbf{20.07} & \textbf{36.66} & 14.64 & \textbf{27.75} \\
\hline
\end{tabular}
\end{table}

In Table~\ref{tab:tablea}, we first compare the two baselines and observe that adding synthetic data to real data is already beneficial when using only the text and speakers in the real dataset. Here, we observe a significant reduction in CER and WER relative on CV test set by 7\% and 6\% over training on the real data.
By considering the case when the sentences are randomly selected (random), it can be observed that augmenting 50~h of real data with 50~h or 100~h of TTS data reduces the WER by 8\% and 11\% relative on CV test set compared to training on real data only. Similarly, augmenting 100~h of real data with 50~h or 100~h of TTS data reduces the WER by 6\% and 9\% relative on CV test set. Similar reductions are achieved in terms of the CER. This shows that TTS based data augmentation can be effective for both small-sized and medium-sized data with greater performance gains obtained when starting from a smaller real dataset or when augmenting with more synthetic data.

Looking at the results for the different sentence selection methods, we observe a statistically significant relative improvement in CER and WER using the natural selection method over other selection methods. This is especially visible when the number of sentences selected is small, i.e., 50~h. The CER and WER then drop by 4\% and 3\% relative to random selection on CV test set. By contrast, the uniform selection method does not yield significantly better results than random selection. As \cite{wu2007data} used a combination of word-, character-, and phoneme-level distributions in their experiments, it is difficult to directly compare their increase in ASR performance using a uniform distribution to ours. Also, an HMM-GMM-based model was used in their experiments which does not have the same behaviour as deep-learning-based models.

When the results are compared to WER on LibriSpeech LS-C and LS-O, we observe some differing patterns: ASR models trained with sentences from random selection have similar WER to those trained on sentences selected using uniform selection. Additionally, ASR models trained on randomly selected sentences also produce similar WER to ASR models trained on the sentences selected using natural selection when the real data size is 100~h. These differences might indicate a bias in the distribution of sentences in both datasets.

In conclusion, although adding synthetic data to the training dataset can reduce the CER and WER of the ASR model, considering the distribution of the text when selecting the sentences for generating the training utterances will further improve its performance.

\begin{figure}
\centering
\subfloat[natural distribution]{\includegraphics[width=0.48\textwidth,trim=15 0 0 0
,clip]{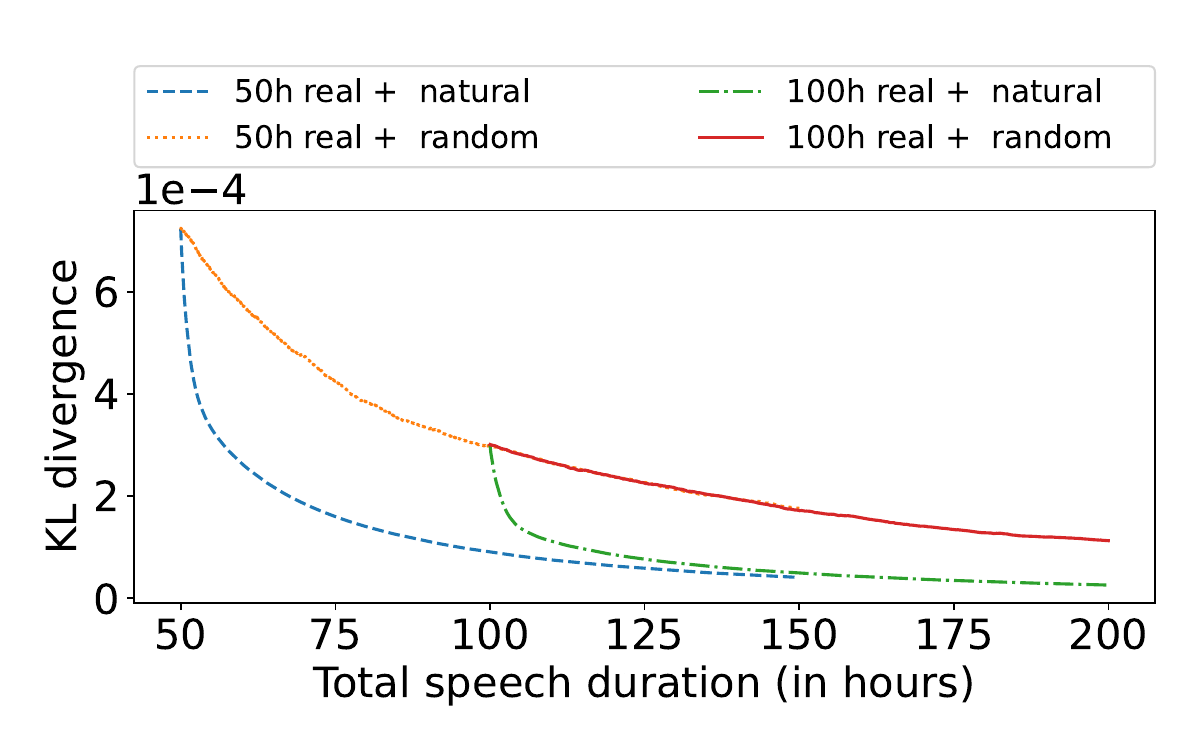}\label{fig:fig_natural}}\hskip1ex
\subfloat[uniform distribution]{\includegraphics[width=0.48\textwidth,trim=15 0 0 0,clip]{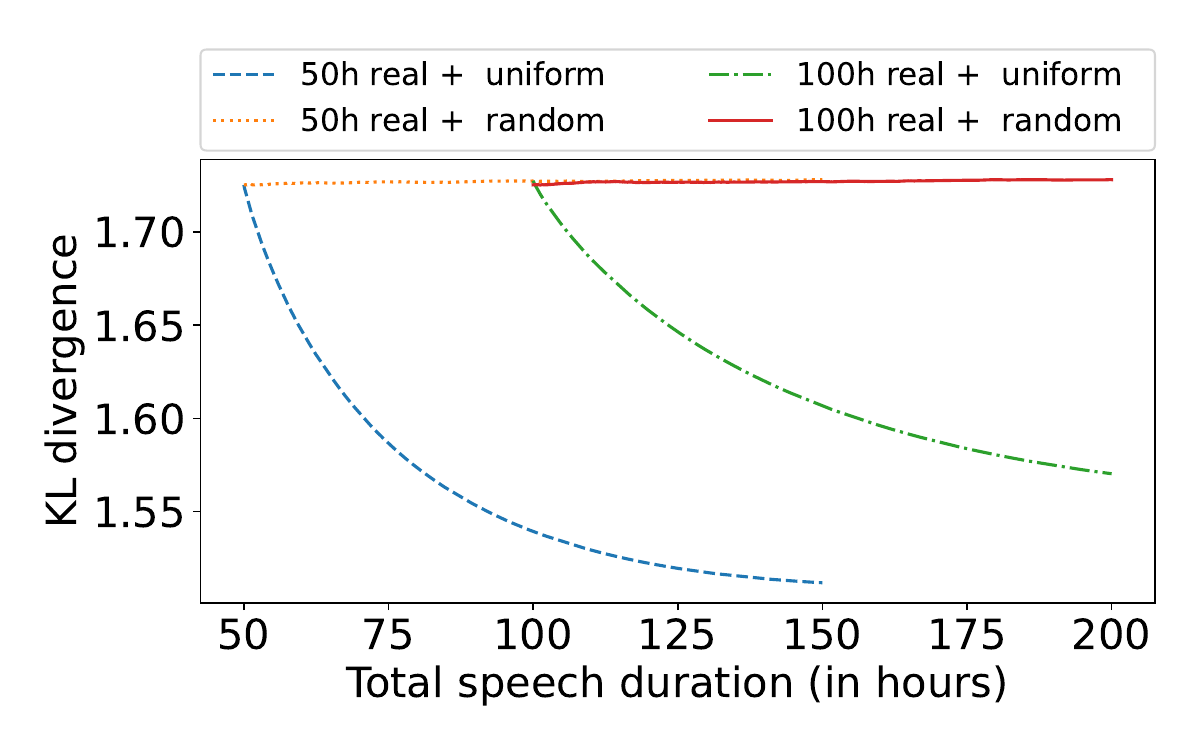}\label{fig:fig_uniform}}
\caption{KL divergence between natural/uniform distribution of texts and combination of real speech texts and newly selected 50~h / 100~h equivalent of TTS data.}
\label{fig:fig_kl}
\end{figure}

To better understand the effect of the different sentence selection methods, we plotted in Figure~\ref{fig:fig_kl} the KL divergence between the distribution of di-phonemes in the set of CV training sentences and the distribution of di-phonemes in the combined set of real sentences and the subsequently added synthetic sentences for each of the selection methods considered compared to random selection. From Figure~\ref{fig:fig_natural}, it can be observed that the natural selection method produces the lowest divergence from the CV data. Although the random selection method reduces the KL divergence, one can extrapolate from Figure~\ref{fig:fig_natural} that it will take twice more data to achieve a similar KL divergence.
Similarly, using the uniform distribution reduces the KL divergence to some extent in Figure~\ref{fig:fig_uniform} but seems to saturate quickly without any further reduction. Additionally, the KL divergence becomes flat when the distribution of randomly selected text is compared to the uniform distribution indicating that the randomly selected data does not closely follow a uniform distribution of di-phonemes.

\subsection{Increasing speaker diversity}
The natural way of acquiring more speakers when there are few speakers in a dataset is to collect more data with more speakers. However, since TTS/VC models are capable of generating data when conditioned on arbitrary speakers, this approach may be used to increase the diversity of speakers for ASR data augmentation.

\subsubsection{Methodology}
\label{speaker-diversity}
Given the set of speakers $S = \{s_1, s_2, s_3, ..., s_n\}$ from the real ASR training dataset and a set of $T$ unseen speakers, where $S\ \cap\ T = \emptyset$, the goal is to select the most informative speakers $R \subset T$ that will improve the ASR performance when used to synthesise additional training data. To achieve this, we explore a selection method that considers the distance of speakers to other speakers in their embedding space.

To maximise the speaker diversity, we successively select speakers in $T$ according to their distance to the real speakers and the already selected speakers.
Using a distance metric, we first compute the distance between each unselected speaker in $T$ and the closest speaker in $S\cup R$.
Then, we select the speaker corresponding to a given distance quantile: the closest speaker to any of the speakers in $S \cup R$ is denoted as \emph{minmin}, the speaker with the median distance is denoted as \emph{medmin}, and the farthest speaker is denoted as \emph{maxmin}. The minmin based speaker selection is meant to avoid outliers while the medmin and maxmin based speaker selection are meant to select more novel speakers.
The iterative algorithm for \emph{maxmin} based speaker selection is described in Algorithm~\ref{alg:alg_spk_select}. Changing arg~max to arg~min or arg~median in Algorithm~\ref{alg:alg_spk_select} changes the algorithm to the \emph{minmin} and \emph{medmin} speaker selection method.

\begin{algorithm}[!ht]
\caption{Speaker selection algorithm using maxmin}\label{alg:alg_spk_select}
\begin{algorithmic}[1]
\Require $S$, $T$, $n$
\Procedure{SelectSpeaker}{$S, T, n$}       \Comment{$n$ is the desired number of speakers}
    \State Initialise $R = \emptyset,\ K = \emptyset$
    \While{$|R| < n$}
        \For{$t$ in $T$}
            \State $d_{t, s} = \min_{s \in S \cup R} \text{dist}(t,s)$
            \State $K := K \cup \{d_{t, s}\}$
        \EndFor
        \State $s = \text{arg}\max_{k \in K} k$
        \State $R:= R \cup \{s\}, T:= T \setminus \{s\}, K = \emptyset$
    \EndWhile
    \State return $R$
\EndProcedure
\end{algorithmic}
\end{algorithm}

Concretely, we compute the cosine distance between the embeddings $e_A$ and $e_B$ of speaker $A$ and speaker $B$ as
\begin{equation}
\label{cosine}
\text{dist}(e_A, e_B) = 1 - \dfrac{e_A . e_B}{\|e_A\| \, \|e_B\|}.
\end{equation}
$S$ is the set of speakers in the real training set and $T$ are the speakers disjoint from $S$ in the CV training dataset. 
Speaker embeddings are used to condition the Glow-TTS model to synthesise 50~h of synthetic data. We iterate over the set of speaker embeddings available during TTS/VC inference to cover all speakers while using 50~h worth of randomly selected sentences. 

In addition to comparing the speaker selection methods, we iteratively increase the number of selected speakers by a factor of 2 to determine the number of synthetic speakers required to significantly improve the performance of the ASR model. Here, we increase the TTS and VC dataset size to 100~h to have enough samples per speaker. The speakers are randomly selected from $T$ with each smaller set being a subset of the larger set.

\subsubsection{Results}
We compare the three speaker selection methods (\emph{maxmin}, \emph{medmin} and \emph{minmin}) to \emph{random} speaker selection from the unseen speaker set. As a baseline, we also provide results for the ASR models trained on augmented data from \emph{seen} speakers only, where the seen speakers are the 2{,}457 speakers in the real ASR training set.

Table~\ref{tab:table2_1} shows the CER and WER of CTD ASR models on the CV test set when trained on a combination of 50~h of real and 50~h of TTS data that has been generated using speakers from the different speaker selection methods. Firstly, the results show a relative reduction on CV test set in CER and WER of 3\% and 2\% of the ASR model trained on a combination of real data and TTS data generated using randomly selected unseen speakers compared to TTS data generated using only the speakers in the real ASR training set (seen). This indicates that ASR performance can be improved by randomly adding more speakers. Secondly, we see a statistically significant difference in the ASR performance between medmin and other sentence selection methods in terms of CER. In comparison to LS-C and LS-O, TD models trained on random speakers produced the lowest WERs.  This may be due to the bias in speakers in the LibriSpeech dataset to native speakers reading audiobooks than the large speaker distribution in CV.

\begin{table}[!ht]
\caption[]{\%~CER and \%~WER of CTD ASR models trained using real data and data synthesised using speakers selected randomly or by using the minmin, medmin, or maxmin speaker selection criterion.}\label{tab:table2_1}
\centering
\begin{tabular}{ l  l | l  l | l  l }
\hline
\multirow{2}*{\makecell{Speaker \\ selection method}} & \multirow{2}*{Data} & \multicolumn{2}{c|}{CV} & LS-C & LS-O \\
\cline{3-6}
 &  & CER & WER & WER & WER \\
\hline
Seen & 50r\_50tts & 23.28 & 42.79 & 18.70 & 35.19 \\ \hline
Random & \multirow{4}{*}{50r\_50tts} & 22.57 & 42.02 & \textbf{18.40} & \textbf{34.15} \\
Minmin &  & 22.55 & \textbf{42.15} & \textbf{18.49} & 34.94 \\
Medmin &  & \textbf{22.23} & \textbf{41.65} & 18.78 & 34.89 \\
Maxmin &  & 22.47 & \textbf{41.79} & 19.11 & 35.25 \\
\hline
\end{tabular}
\end{table}

\begin{figure*}[!hbtp]
\centering
\captionsetup[subfloat]{labelfont=scriptsize,textfont=scriptsize}
\subfloat[Random speaker selection]{\includegraphics[width=1.8in,trim=15 0 0 0,clip]{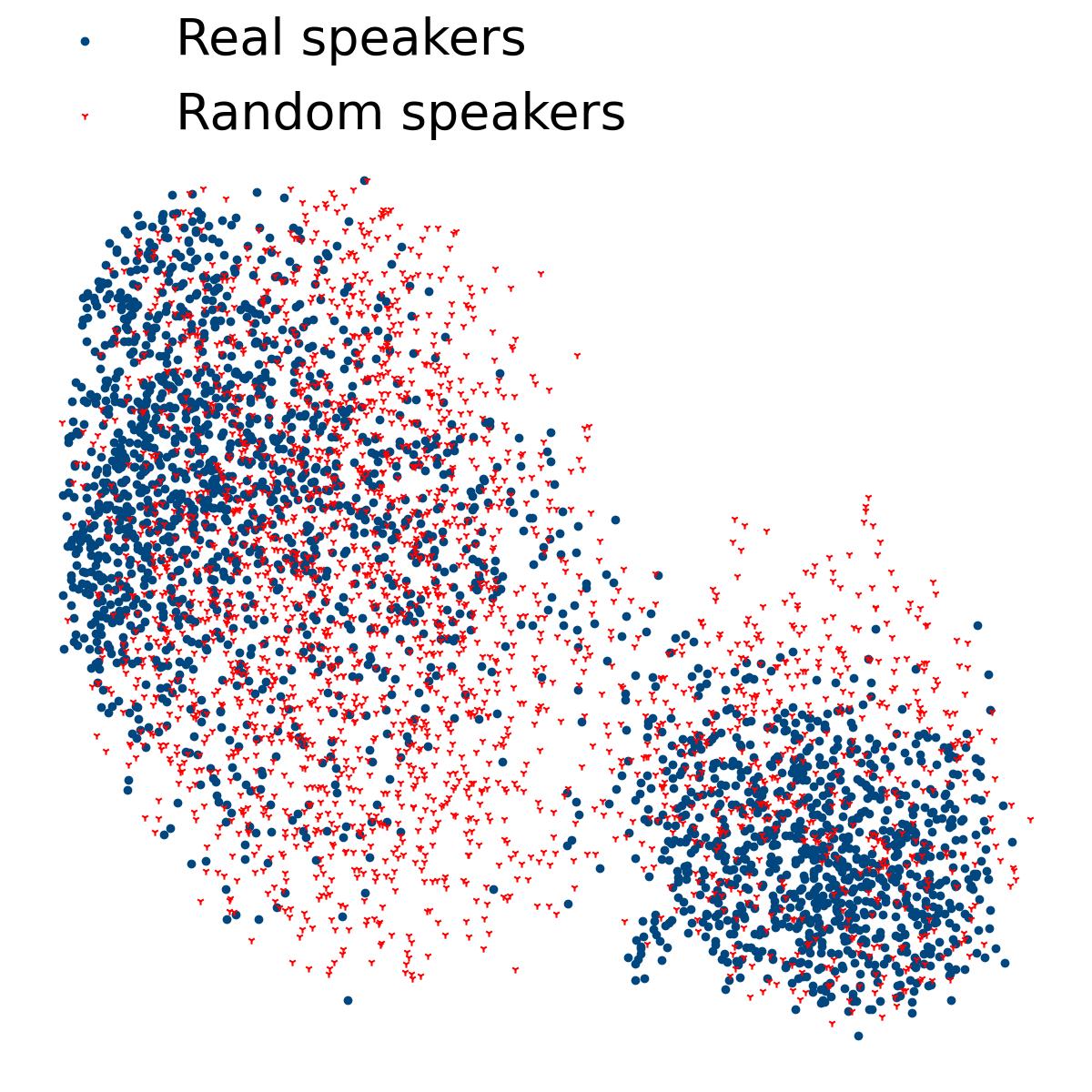}\label{fig:fig_rand}}
\subfloat[Minmin speaker selection]{\includegraphics[width=1.8in,trim=15 0 0 0,clip]{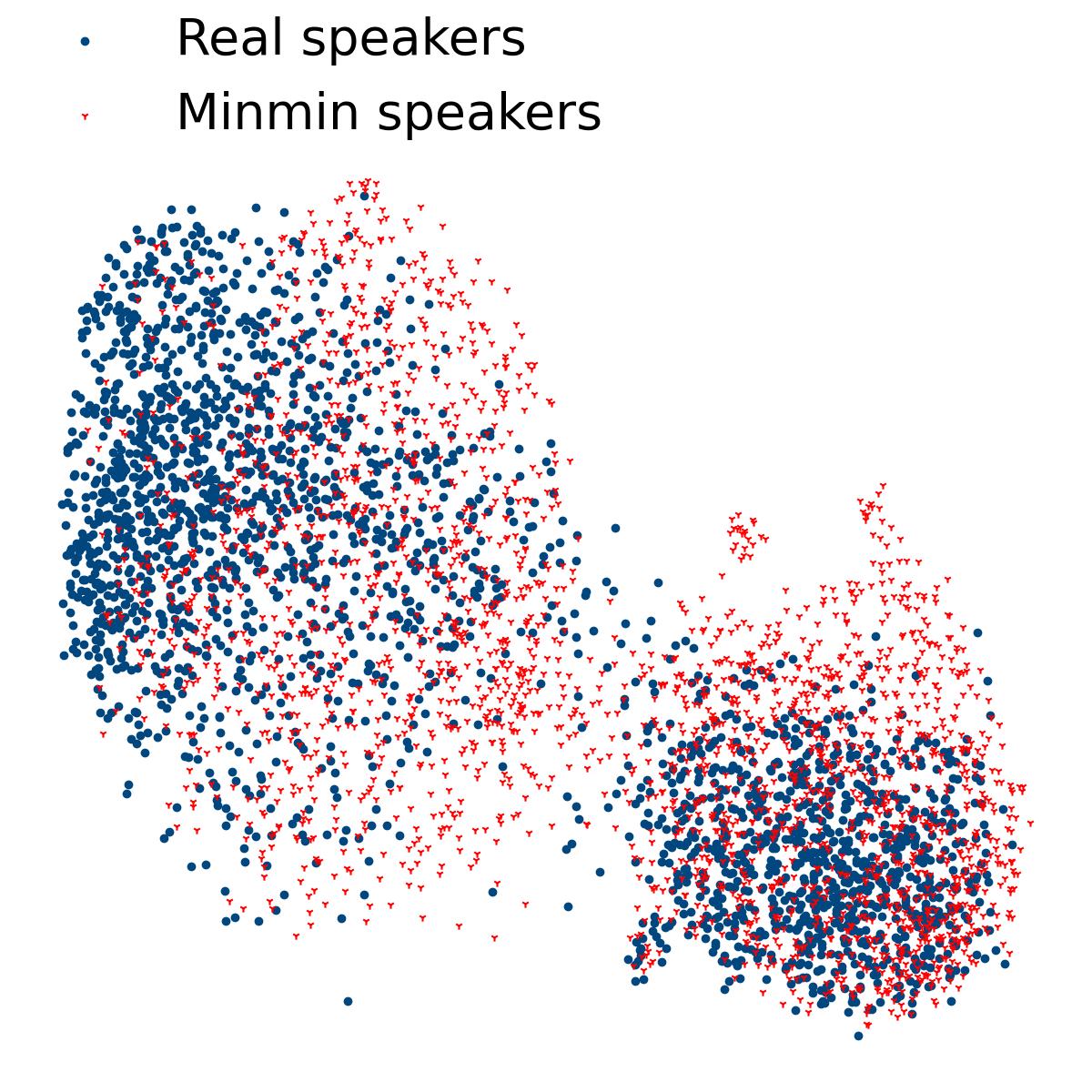}\label{fig:fig_min_min}}
\subfloat[Medmin speaker selection]{\includegraphics[width=1.8in,trim=15 0 0 0,clip]{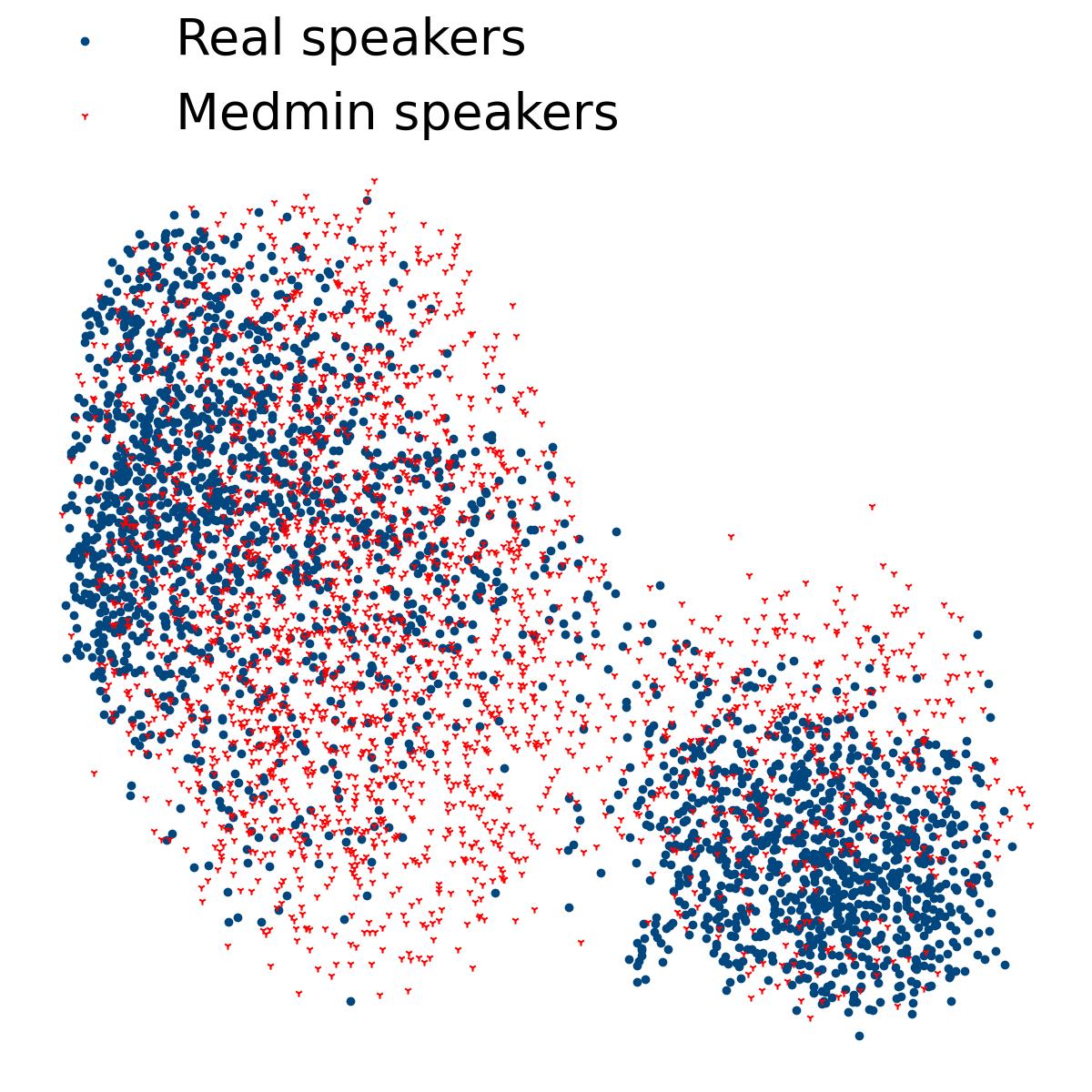}\label{fig:fig_med_min}}
\subfloat[Maxmin speaker selection]{\includegraphics[width=1.8in,trim=15 0 0 0,clip]{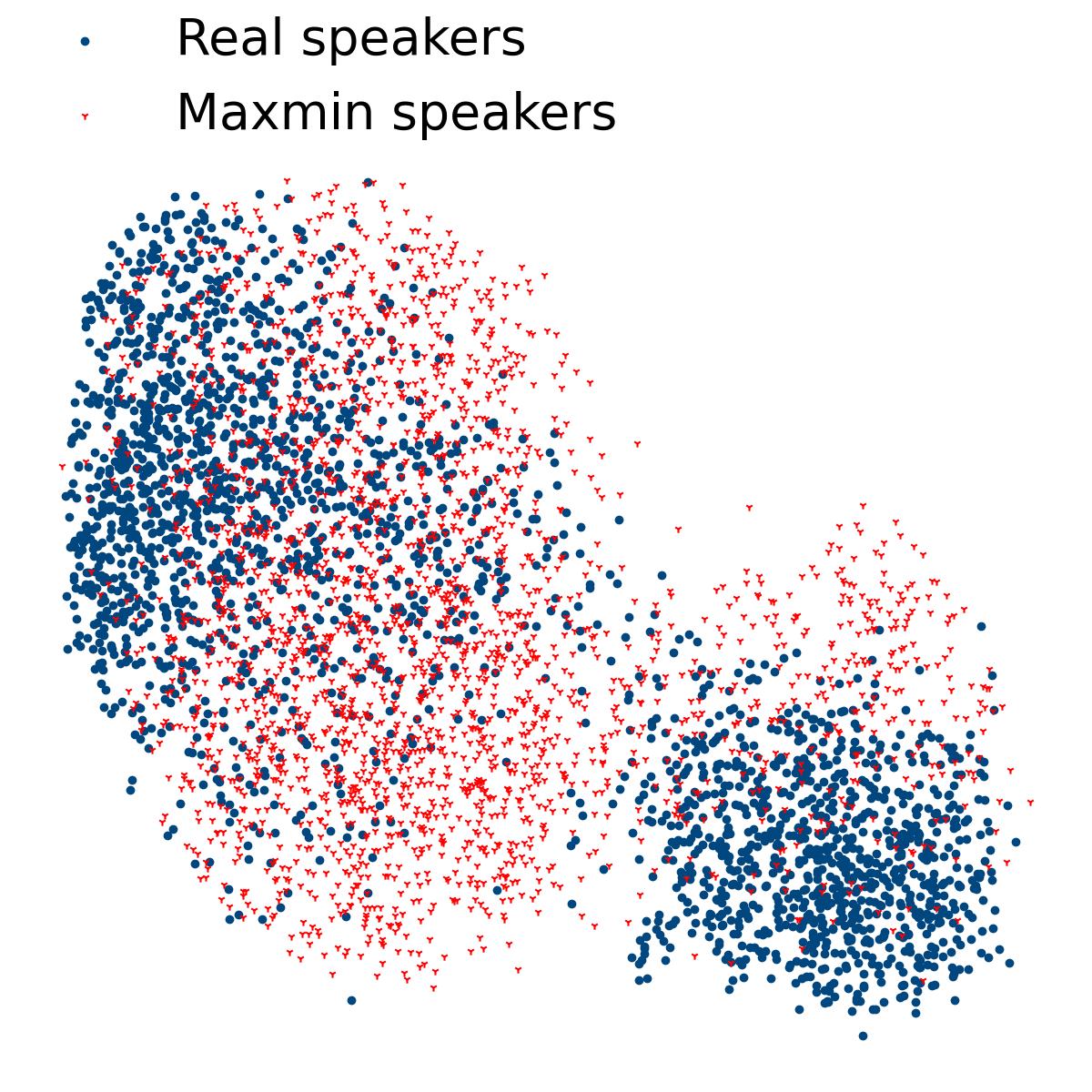}\label{fig:fig_max_min}}
\caption{UMAP plot of speaker embeddings of real data speakers and speakers selected using the speaker selection algorithm.}
\label{fig:fig_spk_select}
\end{figure*}

We plot the speaker embeddings of the real data and those of the selected speakers in Figure~\ref{fig:fig_spk_select} using UMAP \cite{mcinnes2018umap}. UMAP provides a two-dimensional representation of $N$-dimensional embeddings, to reflect the distance between the embeddings as seen in Figure~\ref{fig:fig_min_min}. Each of the plots visibly has two main clusters indicating male and female groups. From these plots, we observe that the minmin selection criterion truly selects speakers around the real speakers and therefore only fills up holes in the speaker distribution. For this reason, its plot is denser than others which can be visually seen from the right side of the plot. Visually, the medmin and maxmin selection criteria produce a wider spread of newly selected speakers and therefore expanded the original speaker distribution. Visually, the randomly selected speakers also had a similar effect as medmin and maxmin on the speaker distribution.

In Table~\ref{tab:table3_1} where the CTD ASR systems are trained on a combination of real data and TTS data generated with an increasing number of speakers, we similarly saw a significant relative decrease in both CER and WER of the ASR models by 2\% on CV test set. Increasing the number of speakers above 2{,}457 does not yield an improvement in the ASR results. We observe that there is a trade-off between increasing the number of speakers in the dataset to the number of samples per speaker. The results seem to indicate that having more samples per speaker may also be as important as adding new speakers. On the LS-C and LS-O test sets, increasing the number of speakers to cover a larger speaker distribution significantly improves the CTD ASR performance when adding TTS data.

\begin{table}[!ht]
\caption{\%~CER and \%~WER of CTD ASR models trained on a combination of real data and TTS generated data when the TTS model is conditioned on different numbers of speakers. Speakers are randomly selected from the candidate set of ASR unseen speakers.}\label{tab:table3_1}
\centering

\begin{tabular}{ l  l | l  l | l  l }
\hline
\multirow{2}{*}{K = 2{,}457 speakers} & \multirow{2}{*}{Data} & \multicolumn{2}{c|}{CV} & LS-C & LS-O \\ \cline{3-6}
 &  & CER & WER & WER & WER \\
\hline
baseline & 50r\_100tts & 22.23 & 40.98 & 16.52 & 31.92 \\
\hline \hline

+\ $K$ 
 & \multirow{4}{*}{50r\_100tts} & \textbf{21.84} & \textbf{40.21} & 16.64 & 32.56 \\

+\ $2K$ 
 &  & 22.39 & 40.92 & 16.91 & 33.21 \\

+\ $4K$ 
 &  & 22.34 & 40.95 & 16.93 & 33.03 \\

+\ $8K$ 
 &  & 22.30 & 41.02 & \textbf{16.34} & \textbf{32.42} \\ \hline

\end{tabular}
\end{table}

\begin{table}[!ht]
\caption{\%~CER and \%~WER of CTD ASR models trained on a combination of real data and VC-generated data when the VC model is conditioned on different numbers of speakers. Target speakers randomly selected from the candidate set of ASR unseen speakers.}\label{tab:table3_2}
\centering

\begin{tabular}{ l  l | l  l | l  l }
\hline
\multirow{2}{*}{K = 2457 speakers} & \multirow{2}{*}{Data} & \multicolumn{2}{c|}{CV} & LS-C & LS-O \\
\cline{3-6}
 &  & CER & WER & WER & WER \\
\hline
baseline & 50r\_100vc & 24.79 & 46.77 & 23.50 & 40.91 \\
\hline \hline
+\ $K$
 & \multirow{4}{*}{50r\_100vc} & \textbf{24.60} & \textbf{46.04} & 23.85 & 41.34 \\
+\ $2K$
 &  & \textbf{24.33} & \textbf{46.09} & \textbf{22.82} & \textbf{39.65} \\
+\ $4K$
 &  & \textbf{24.61} & \textbf{46.03} & \textbf{22.71} & \textbf{39.63} \\
+\ $8K$
 &  & \textbf{24.41} & \textbf{46.09} & 23.38 & 40.69 \\ \hline

\end{tabular}
\end{table}
We observe similar results for ASR models trained on real and VC speech as shown in Table~\ref{tab:table3_2}. For these experiments, VC is performed on each real utterance by iterating through the specified number of random ASR unseen speakers until a VC data size of 100~h is reached. Here, we see a significant drop in the ASR's CER and WER by 1\% and 2\% relative on CV test set when 2{,}457 TTS unseen speakers are used for VC. Increasing the number of speakers beyond 2{,}457 does not yield any statistical improvement in the results. Compared to LS-C and LS-O test sets, we observe a relative reduction in WERs until $4K$ speakers with additional speakers not reducing the WERs further.
In addition, VC experiments do not produce similar (or better ASR results) to TTS experiments perhaps due to the lack of augmentation of phonetic diversity.

\begin{figure*}[!hbtp]
\centering
\captionsetup[subfloat]{labelfont=scriptsize,textfont=scriptsize}
\subfloat[K speakers]{\includegraphics[width=1.8in,trim=15 0 0 0,clip]{files/rand_spks.jpg}\label{fig:fig_rand_k}}
\subfloat[2K speakers]{\includegraphics[width=1.8in,trim=15 0 0 0,clip]{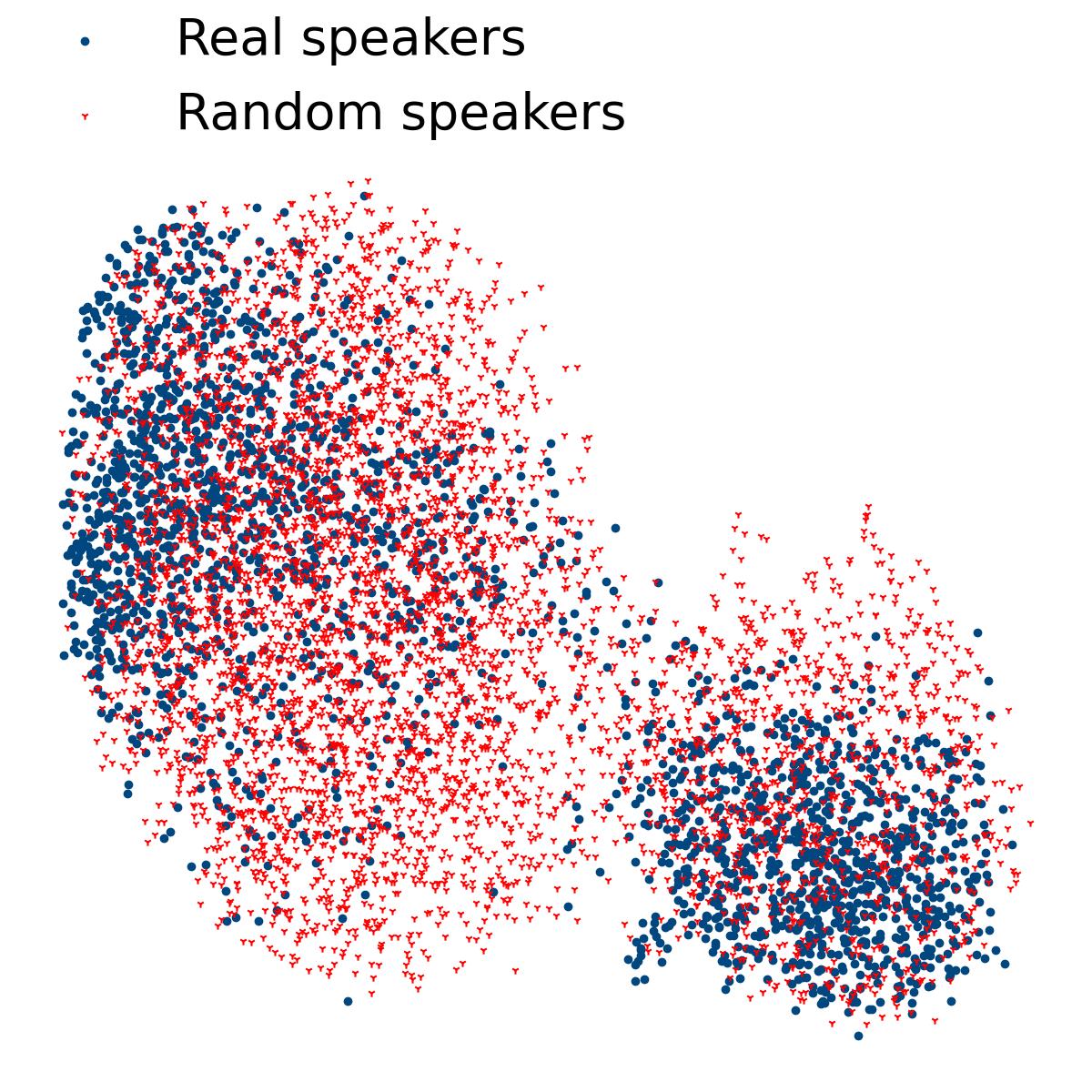}\label{fig:fig_rand_2k}}
\subfloat[4K speakers]{\includegraphics[width=1.8in,trim=15 0 0 0,clip]{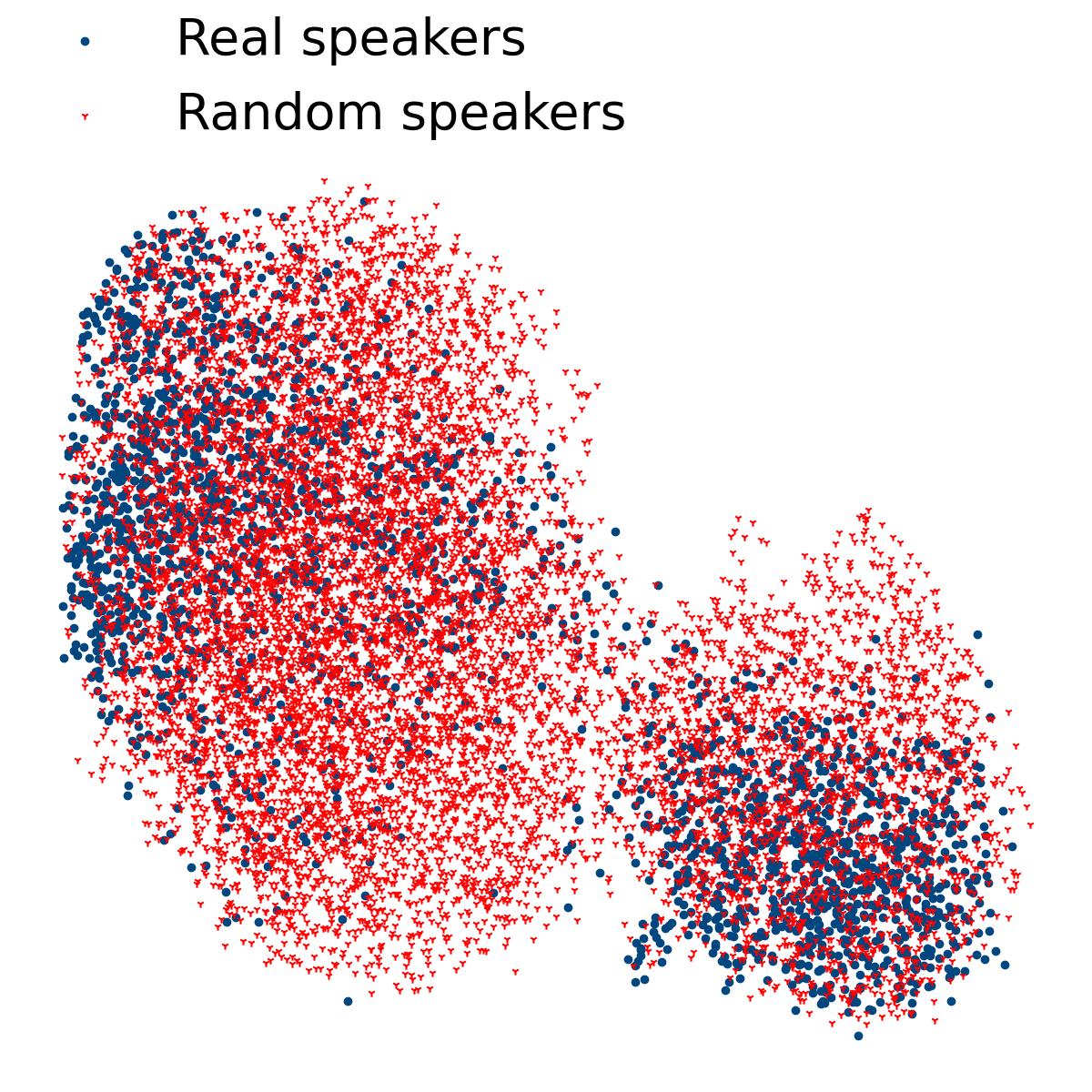}\label{fig:fig_rand_4k}}
\subfloat[8K speakers]{\includegraphics[width=1.8in,trim=15 0 0 0,clip]{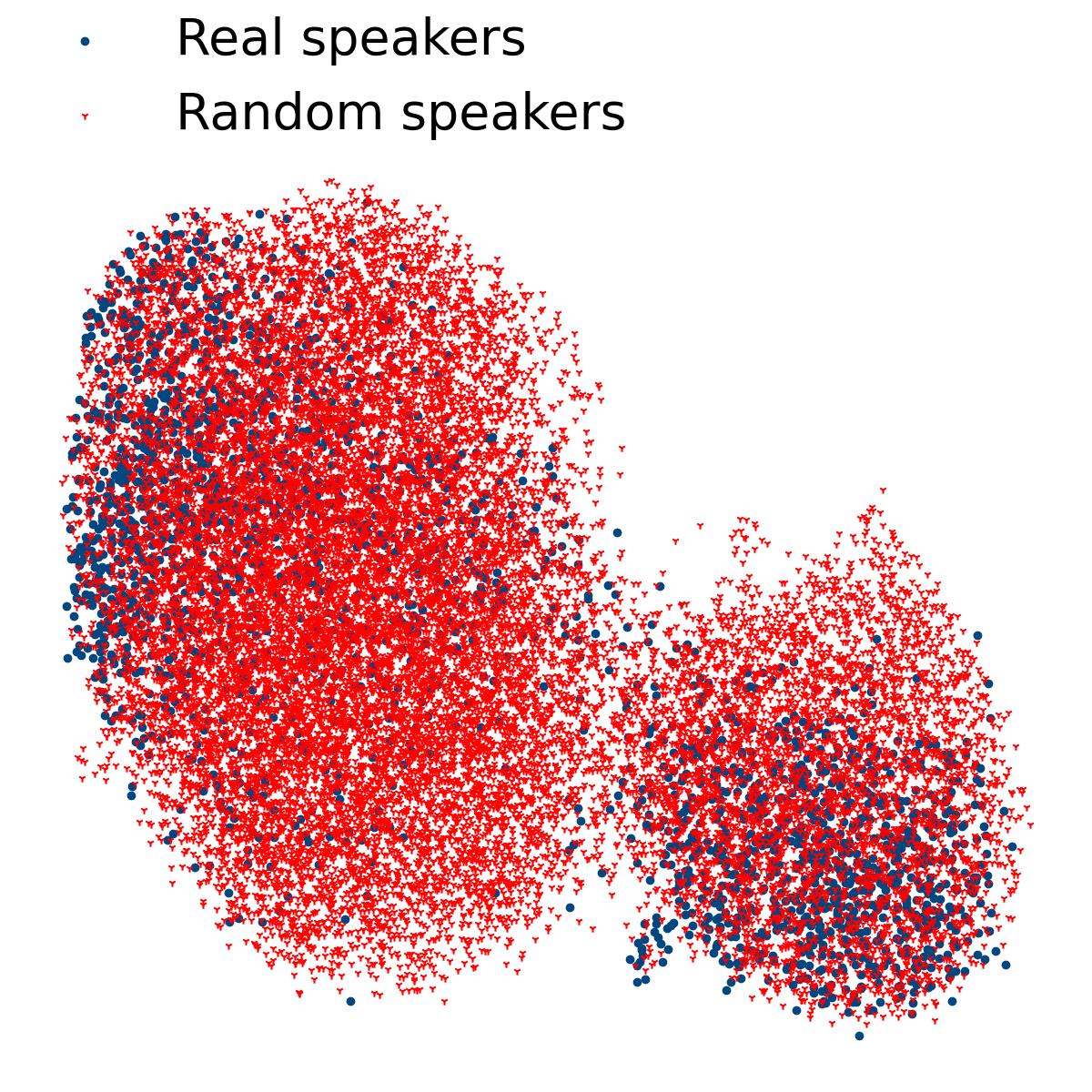}\label{fig:fig_rand_8k}}
\caption{UMAP plot of speaker embeddings of real data speakers and $K$, $2K$, $4K$ or $8K$ randomly selected speakers not in the real dataset with $K=2{,}457$.}
\label{fig:fig_random_spk_select}
\end{figure*}

In Figure~\ref{fig:fig_random_spk_select}, we plot the speaker embeddings of the real data with the speaker embeddings of $K$, $2K$, $4K$ or $8K$ randomly selected speakers that are not in the real ASR training set. The distribution of speakers became denser as more speakers were added to the real speaker set. 

\subsection{Increasing phoneme duration diversity}
\label{duration_augmentation}

A well-known technique for augmenting an ASR dataset is to either increase or decrease the speaking rate of the utterances through speed perturbation \cite{ko2015audio}. Speed perturbation changes the duration and the spectral content of the utterances but does not affect the relative duration of phonemes within each utterance. 
As a data augmentation strategy, we evaluate the impact of modifying relative phoneme durations instead by using a duration predictor to predict diverse phoneme durations in the TTS/VC model. We also consider time stretching to modify the duration of the utterances without affecting their spectral content.

\subsubsection{Methodology}
Our method involves only augmenting the phoneme duration while keeping the speaker characteristics and phonetic content of the utterances. Therefore, the duration augmentation method requires that we provide two different speaker embeddings to the TTS/VC model, the speaker embedding of the original real speech sample and another randomly chosen speaker embedding from the pool of real speaker embeddings. The original or target speaker embedding is used to condition the decoder during generation and for MAS alignment of the original utterance (in the case of VC) while the duration predictor is conditioned on the randomly chosen speaker embedding to generate the phoneme duration. 

In the case of VC, the aligned phoneme duration of the original real speaker and the predicted phoneme duration of a randomly chosen speaker are also required. 
Since the length of some phonemes for the original and the randomly chosen speaker will differ, we either duplicate or drop frames in the latent representation to match the newly predicted duration. Making the changes in the latent representation does not utterly degrade the naturalness of the synthesised utterances, except for some noticeable co-articulation errors. We therefore set a maximum threshold of five frames that can be dropped or added for any phoneme duration. Speed perturbation can also be performed by resampling the original speech signal, however this affects the spectral content. Here, since we want to evaluate only the duration, we apply time-stretching as implemented in the TimeStretch class in the Audiomentations library.\footnote{\label{refnote}\url{https://github.com/iver56/audiomentations.git}}

In addition to the main experiments, we include four baselines which combine the real data with the following derived data:
\begin{itemize}
    \item \emph{50r\_50rt}: time-stretching the real data to create an augmented time-stretched version of the same size,
    \item \emph{50r\_50rto}: time-stretching the real data on the fly with a probability of 0.5 on every dataset iteration,
    \item \emph{50r\_50rtts}: using the original real speaker embedding for predicting the durations for TTS augmentation, and
    \item \emph{50r\_50rvc}: using a random speaker embedding for VC without changing the duration of the original utterance.
\end{itemize}
We evaluate the duration-based data augmentation with 50~h of real speech. Here, we denote the datasets augmented with the synthetic data using our method of generating diverse durations as $50r\_50tts$ and $50r\_50vc$.

We set the minimum stretching rate and maximum stretching rate to $0.9$ and $1.1$, following previous data augmentation studies by \cite{ko2015audio} and \cite{rossenbach2023relevance}.  
TTS/VC-based utterance generation is done using Glow-TTS and GlowTTS-STD. As previously described in Section~\ref{tts_models}, GlowTTS-STD uses a generative flow-based duration predictor to generate phoneme durations. We iterate over the set of real speakers during TTS/VC generation to cover all speakers.

\subsubsection{Results}

Table~\ref{tab:table4_1} shows the results of these experiments. Beginning with the CTD ASR models trained with time-stretching, generating the time-stretched speech samples on the fly decreases the CER and WER by 4\% and 5\% relative on CV test set compared to generating a 50-h time-stretched version of the real data. We also observe that adding TTS data to the real data provides a greater improvement to the performance of the ASR systems. In particular, naively regenerating the real utterance's text using Glow-TTS with the same real speaker's embedding reduces the CER and WER of the ASR system by 4\% and 3\% relative on CV test set over applying time-stretching to the real data with a probability of $0.5$. Applying a randomly selected speaker to drive the duration predictor in Glow-TTS for duration diversity further reduces the WER by 1\% relative on CV test set, although we do not see a significant difference in the CER. While using the GlowTTS-STD model provides a significant improvement over the time-stretching baselines, it does not improve the ASR system relative to Glow-TTS. Although our speaker augmentation method reduces the WERs on LS-C and LS-O relative to classical augmentation methods. Using random speakers for augmentation did not yield any significant improvements. 
Lastly, CTD ASR models using VC-generated utterances for data augmentation had the lowest performance.
\begin{table}[!ht] \caption{\%~CER and \%~WER on the CV test set of CTD ASR models trained on a combination of 50~h real and synthetic data where different duration augmentation methods have been applied to increase duration diversity.  
} \label{tab:table4_1}
\centering

\begin{tabular}{ l  l | c  c | c  c }
\hline
\multirow{2}{*}{Model} & \multirow{2}{*}{Data} & \multicolumn{2}{c|}{CV} & LS-C & LS-O \\
\cline{3-6}
 &  & CER & WER & WER & WER \\
\hline
\multicolumn{6}{l}{Baselines} \\
\hline
- & 50r & 25.01 & 46.30 & 21.18 & 37.74 \\ 
- & 50r\_50rt & 25.09 & 47.28 & 23.18 & 38.97 \\
- & 50r\_50rto & 24.04 & 44.83 & 21.69 & 37.59 \\
GlowTTS & 50r\_50rtts & \textbf{23.19} & 43.44 & \textbf{20.32} & \textbf{35.81} \\
GlowTTS & 50r\_50rvc & 25.02 & 46.97 & 23.01 & 40.58 \\
\hline
\multicolumn{6}{l}{Text-to-speech} \\
\hline
GlowTTS & \multirow{2}{*}{50r\_50tts} & \textbf{23.28} & \textbf{42.95} & \textbf{20.46} & 36.38 \\
GlowTTS-STDP &  & 23.95 & 43.31 & 21.97 & 36.96 \\
\hline
\multicolumn{6}{l}{Voice conversion} \\
\hline
GlowTTS & \multirow{2}{*}{50r\_50vc} & 24.44 & 45.06 & 22.39 & 38.86 \\
GlowTTS-STD &  & 25.85 & 47.68 & 23.48 & 40.53 \\
\hline

\end{tabular}
\end{table}

\subsection{Increasing pitch diversity}

Pitch modification has been applied for data augmentation in some ASR studies, e.g., by \cite{casanova23_interspeech}. Besides, classical speed perturbation based data augmentation results in a modification of the pitch. For this reason, we examine the impact of pitch diversity in ASR by comparing augmentation using TTS/VC models to classical pitch modification methods.

\subsubsection{Methodology}

Similar to what was described in Section~\ref{duration_augmentation} for generating diverse phoneme durations, here we only want to change the pitch without changing the phonetic or prosodic attributes of the utterance. As such, we provide two different speaker embeddings to the TTS/VC model: the speaker embedding of the original or target real speech sample and another randomly chosen speaker embedding from the pool of real speaker embeddings. The original speaker embedding is used to generate phoneme durations and to condition the decoder during Mel-spectrogram generation or inversion (in the case of VC) while the randomly chosen speaker embedding is used to condition the pitch predictor for generating per-frame log-F0 values. 

We also included two baselines which combine real data with the following derived data:
\begin{itemize}
    \item \emph{50r\_50rp}: pitch-shifting the real data to create an augmented version of the same size,
    \item \emph{50r\_50rpo}: pitch-shifting the real data on the fly with a probability of $0.5$ on every dataset iteration.
\end{itemize}
Along with the baselines, we train ASR models using the real data combined with the following data generated either by TTS or VC:
\begin{itemize}
    \item \emph{50r\_50tts\_osp}: TTS data generated by GlowTTS-STDP using the original real speaker embedding for pitch prediction,
    \item \emph{50r\_50tts\_rsp}: TTS data generated by GlowTTS-STDP using a random speaker selected from the real speaker set,
    \item \emph{50r\_50vc\_rsp}: VC data converted by GlowTTS-STDP by providing predicted log-F0 values of a random speaker. Here, we use the original speaker embedding for conditioning the vocoder.
\end{itemize} 

For pitch shifting, a value is randomly chosen in the range of -2 to +2 semitones.
The pitch-shifting augmentation is performed using the PitchShift class in the Audiomentations library.\footref{refnote} TTS/VC-based pitch augmentation is done using GlowTTS-STDP as it is the only model that uses external pitch information or generates log-F0 values. Also, we iterate over the set of real speakers during TTS/VC generation to cover all speakers.

\subsubsection{Results}

Table~\ref{tab:table5_1} shows the results of these experiments.
Firstly, we observe that applying pitch shifting on the fly with a $0.5$ probability negatively impacts the ASR results and performs worse than generating a 50~h pitch-shifted version of the real data. This indicates that a large amount of pitch-shifting can have an unintended negative effect on ASR performance.
Secondly, we observe that using a random speaker embedding for pitch prediction reduces both the CER and WER on CV test set by 1\% relative over using the original real speaker embedding for pitch prediction. Although this shows that the model benefits from pitch diversity, it is still worse than ASR models using a standard Glow-TTS without pitch modifications. Additionally, comparing Table~\ref{tab:table4_1} to Table~\ref{tab:table5_1} indicates that duration augmentation is more beneficial to ASR training data augmentation than pitch augmentation. We also observe similar conclusions with the LS-C and LS-O test sets. Lastly, as has been the case with other experiments, the ASR model does not benefit from adding pitch-augmented VC data to the real data.

\begin{table}[!ht] 
\caption{\%~CER and \%~WER on the CV test set of CTD ASR models trained on a combination of 50~h real and synthetic data where different pitch augmentation methods have been applied to increase pitch diversity.
} \label{tab:table5_1}
\centering

\begin{tabular}{ l  l | c  c | c  c }
\hline
\multirow{2}{*}{TTS Model} & \multirow{2}{*}{Data} & \multicolumn{2}{c|}{CV} & LS-C & LS-O \\
\cline{3-6}
 &  & CER & WER & WER & WER \\
\hline
- & 50r & 25.01 & 46.30 & \textbf{21.18} & \textbf{37.74} \\
- & 50r\_50rp & 24.96 & 46.66 & 23.13 & 39.93 \\
- & 50r\_50rpo & 26.04 & 48.19 & 25.17 & 42.40 \\
\hline
\multicolumn{6}{l}{Text-to-speech} \\
\hline
\multirow{2}{*}{GlowTTS-STDP} & 50r\_50tts\_osp & 24.30 & 44.59 & 22.41 & \textbf{37.64} \\
 & 50r 50tts\_rsp & \textbf{24.06} & \textbf{44.14} & 22.80 & \textbf{37.99} \\
\hline
\multicolumn{6}{l}{Voice conversion} \\
\hline
GlowTTS-STDP & 50r\_50vc\_rsp & 25.48 & 47.33 & 22.75 & 39.84 \\
\hline
\end{tabular}
\end{table}

\subsection{Matching the environmental conditions of real speech}

In the previous experiments, all the ASR models were trained on a combination of real and synthetic data without considering the difference in environmental conditions of the combined datasets. Therefore, in this section, we consider the impact of synthetic data augmented with ambient noise and varying room characteristics on the performance of the ASR systems.

\subsubsection{Methodology}

To simulate a more noisy dataset, additive noise is randomly added to the synthesised utterances with a random signal-to-noise ratio (SNR) between 0~dB and 15~dB. Similarly, to simulate reverberation in speech, the synthetic utterances are randomly convolved with room impulse response (RIR) filters. We modified the TTS/VC-generated utterances with a probability of 0.25 or 0.5. 
Finally, we performed experiments covering both augmentation methods.

Concretely, in each training set batch, if a given synthetic utterance is selected for reverberation augmentation it is convolved with a random simulated RIR and if it is selected for noise augmentation a random noise sample is added to it (or to the reverberated utterance in case it was also selected for reverberation augmentation). The added noises are taken from the 6~h noise subset of the MUSAN corpus \cite{musan2015} which consists of technical and non-technical noises while the simulated RIRs are those provided by \cite{7953152}.

\subsubsection{Results}

Tables~\ref{tab:table6_1} and \ref{tab:table6_2} show the results for ASR models trained with real data and augmented TTS/VC-generated data. Firstly, concerning TTS-generated data in Table~\ref{tab:table6_1}, we observe that training on synthetic data with noise or reverberation improves the performance of the ASR models. With 25\% noise augmentation probability, we observe a 3\% and 5\% relative drop in CER and WER on CV test set over training without noise. Increasing the probability to 50\% further reduces the CER and WER by 7\% and 5\% relative on CV test set compared to training without noise. Secondly, adding reverberation to 25\% of the TTS-generated utterances reduces the CER and WER by 1\% and 2\% relative on CV test set. Increasing the probability to 0.5 does not improve the ASR performance. In general, adding noise or reverberation is better than using only the TTS-generated data without augmentation. Lastly, combining noise and reverberation at different probabilities reduces the CER and WER compared to using either noise or reverberation on CV test set. For LS-C and LS-O test sets, we observe that adding noise helps but adding reverberation does not reduce WERs on the test datasets. This is probably due to the quality of LS-C and LS-O test sets which do not contain similar environmental conditions as CV. 
\begin{table}[!ht]\caption{\%~CER and \%~WER of CTD ASR models trained on a combination of 50~h real and 50~h TTS generated data augmented with noise or RIRs with different probabilities (Prob).}\label{tab:table6_1}
\centering

\begin{tabular}{ l  l | l  l | l  l }
\hline
\multirow{2}{*}{Data} & \multirow{2}{*}{Prob} & \multicolumn{2}{c|}{CV} & LS-C & LS-O \\
\cline{3-5}
 &  & CER & WER & WER & WER \\
\hline
50r\_50tts & 0.0 & 23.19 & 43.44 & 20.32 & 35.81 \\
\hline
\multicolumn{6}{l}{Noise added to TTS-generated data} \\
\hline
\multirow{2}{*}{50r\_50tts} & 0.25 & 22.02 & 42.12 & 20.61 & 36.92 \\
 & 0.5 & \textbf{21.47} & 41.20 & \textbf{19.88} & \textbf{35.20} \\
\hline
\multicolumn{6}{l}{Reverb added to TTS-generated data} \\
\hline
\multirow{2}{*}{50r\_50tts} & 0.25 & 22.92 & 42.62 & 20.82 & 36.28 \\
 & 0.5 & 22.91 & 42.67 & 20.57 & 36.16 \\
\hline
\multicolumn{6}{l}{Noise + Reverb} \\
\hline
\multirow{2}{*}{50r\_50tts} & 0.25 & 21.89 & 41.59 & 21.40 & 36.68 \\
 & 0.5 & \textbf{21.55} & \textbf{40.56} & 21.42 & 36.57 \\ \hline

\end{tabular}
\end{table}

Similarly, concerning VC generated data in Table~\ref{tab:table6_2}, while adding either noise or reverberation reduces the CER and WER of the ASR model, adding noise has a larger impact on ASR performance. In particular, adding noise and reverberation degraded the ASR performance compared to adding noise only. This indicates that VC-generated data mostly benefits from added noise and not from simulating reverberation. 

\begin{table}[!ht]\caption{\%~CER and \%~WER of CTD ASR models trained on a combination of 50~h real and 50~h VC-generated data augmented with noise or RIR at different probabilities (prob).}\label{tab:table6_2}
\centering
\begin{tabular}{ l | l | l  l | l  l }
\hline
\multirow{2}{*}{Data} & \multirow{2}{*}{Prob} & \multicolumn{2}{c|}{CV} & LS-C & LS-O \\
\cline{3-6}
 &  & CER & WER & WER & WER \\
\hline
50r\_50vc & 0.0 & 25.02 & 46.97 & \textbf{23.01} & 40.58 \\
\hline
\multicolumn{6}{l}{Noise added to VC-generated data} \\
\hline
\multirow{2}{*}{50r\_50vc} & 0.25 & 22.97 & 44.28 & \textbf{23.13} & \textbf{39.30} \\
 & 0.5 & \textbf{22.66} & \textbf{43.63} & \textbf{22.97} & \textbf{39.32} \\
\hline
\multicolumn{6}{l}{Reverb added to VC-generated data} \\
\hline
\multirow{2}{*}{50r\_50vc} & 0.25 & 24.43 & 45.88 & 23.40 & 40.13 \\
 & 0.5 & 24.44 & 45.79 & 23.72 & 40.60 \\
\hline
\multicolumn{6}{l}{Noise + Reverb} \\
\hline
\multirow{2}{*}{50r\_50vc} & 0.25 & 23.18 & 44.51 & \textbf{23.26} & \textbf{39.48} \\
 & 0.5 & 22.84 & 43.95 & 24.30 & 41.08 \\
\hline
\end{tabular}

\end{table}

\subsection{Combining all the attributes}
\subsubsection{Methodology}
In a final set of experiments, we combine the augmentation methods explored above that independently reduced the CER and WER of the CTD ASR model on CV test set. Therefore, we select the following methods:
\begin{itemize}
    \item Phonetic diversity: sentences selected using the natural selection strategy.
    \item Speaker diversity: $2{,}457$ unseen speakers selected using the medmin speaker selection criterion.
    \item Duration diversity: augmented phoneme duration using random speaker embeddings from the $2{,}457$ unseen speakers.
    \item Pitch diversity: not included in experiments because TTS/VC utterances generated by the GlowTTS-STDP model did not significantly reduce the WER relative to Glow-TTS.
    \item Matching the environmental conditions of real speech: applied reverberation and noise to TTS generated utterances with 50\% probability on every data iteration.
    \item Model: using Glow-TTS for TTS only since VC-generated utterances do not reduce the WER significantly in comparison to TTS-generated utterances.
\end{itemize}
We iteratively increase the size of the synthetic data by a factor of $2$ until a 400~h threshold to observe the effect of data size. Here we run three sets of experiments: a) using the CTD ASR model b) using a wav2vec2 model c) using a wav2vec model with sentences in the real data only. This last experiment enables us to examine the effect of the other speech attributes when the phonetic content cannot be modified, similar to classical ASR training data augmentation methods.

\subsubsection{Results}

Table~\ref{tab:table7_1} shows the CER and WER of the CTD ASR models trained on the combination of attributes. We have added some baselines to the table for easy comparison. We included a baseline of 50~h and 100~h of real data and a baseline of 50~h of real data that has been time-stretched with a probability of 0.5. We copy the results from their respective tables. Firstly, we observe that combining the data augmentation methods reduces the CER and WER of an ASR model trained on either 50~h or 100~h of real data. It reduces the CER and WER on CV test set of the ASR model trained on the 100~h of real data by 4\% and 1\% relative on CV test set. Here, we believe that noise addition to the synthetic data has contributed largely to this improvement. Increasing the size of the TTS data added to the 50~h real data continues to improve the ASR model until 200~h of data on CV, LS-C and LS-O, indicating a limit to the improvement that can be achieved with the synthetic data when the ASR model is randomly intialised.
\begin{table}[!ht]
\caption{\%~CER and \%~WER of CTD ASR models trained on a combination of 50~h real data and different sizes of TTS generated data combining all the attributes that significantly reduced CER and WER.}\label{tab:table7_1}
\centering
\begin{tabular}{ l | l  l | l  l }
\hline
\multirow{2}{*}{Data} & \multicolumn{2}{c|}{CV} & LS-C & LS-O \\
\cline{2-5}
 & CER & WER & WER & WER \\
\hline
50r & 25.01 & 46.30 & 21.18 & 37.74 \\
100r & 22.10 & 40.81 & 16.14 & 30.98 \\ \hline
50r\_50tts & 21.17 & 40.39 & 18.37 & 34.36 \\
50r\_100tts & 20.05 & 38.32 & 16.76 & \textbf{31.64} \\
50r\_200tts & \textbf{19.70} & \textbf{37.07} & 15.58 & \textbf{31.23} \\
50r\_400tts & \textbf{19.70} & 37.31 & \textbf{15.10} & \textbf{31.23} \\
\hline

\end{tabular}
\end{table}

Additionally, synthetic data augmentation can provide higher benefits in comparison to a classical data augmentation method (time-stretching). Here, we observe a 12\% and 11\% relative reduction in CER and WER on CV test set of CTD ASR model trained on a combination of 50~h of real data and 50~h of diverse synthetic data over an ASR model trained on the 50~h time-stretched data.

In Table~\ref{tab:table7_2} where the wav2vec2 model is finetuned on the generated synthetic data, we observe a significant relative reduction in CER and WER until 400~h of synthetic data. Here we record a relative reduction in WER of the wav2vec2 model of 11\% on the ASR dataset with 50~h real data and 400~h synthetic data over the ASR model trained on 50~h real dataset. A greater reduction in WER on LS-C and LS-O test sets is even observed, a 35\% and 20\% relative reduction on LS-C and LS-O test sets on ASR training data with 50~h real data and 400~h synthetic data over ASR model trained on 50~h real data. This indicates the higher robustness of the self-supervised wav2vec model to synthetic data.

\begin{table}[!ht]\caption{\%~CER and \%~WER of wav2vec2 model trained on a combination of 50~h real data and different sizes of TTS generated data combining all the attributes that significantly reduced CER and WER.}\label{tab:table7_2}
\centering
\begin{tabular}{ l | l  l | l  l }
\hline
\multirow{2}{*}{Data} & \multicolumn{2}{c|}{CV} & LS-C & LS-O \\
\cline{2-5}
 & CER & WER & WER & WER \\
\hline
50r & 18.39 & 41.41 & 19.07 & 30.68 \\
100r & 17.10 & 37.87 & 15.29 & 26.81 \\ \hline
50r\_50tts & 18.41 & 41.46 & 19.07 & 30.67 \\
50r\_100tts & 19.35 & 40.19 & 16.22 & 28.32 \\
50r\_200tts & 18.57 & 38.25 & 13.88 & 25.65 \\
50r\_400tts & \textbf{18.04} & \textbf{36.98} & \textbf{12.41} & \textbf{24.55} \\
\hline
\end{tabular}
\end{table}

Lastly, we provide results of wav2vec2 models trained on different sizes on synthetic data added to 50~h of real data to determine the effect of other speech attributes in TTS augmentation while excluding the phonetic content. The results of this experiment are shown in Table~\ref{tab:table7_3}. First, as previously observed under phonetic augmentation, combining TTS data with real data reduces WER even when then the real data sentences are used for TTS data generation. Increasing the size of the TTS data did not significantly improve ASR model performance. Compared to results in Table~\ref{tab:table7_2}, we can also conclude that phonetic augmentation is an integral part of ASR data augmentation using synthetic data as results are generally worse when the phonetic content of the data is not varied.

\begin{table}[!ht]\caption{\%~CER and \%~WER of wav2vec2 model trained on a combination of 50~h real data and different sizes of TTS generated data where TTS text is the same as those in the real training data and all the other attributes that significantly reduced CER and WER are modified .}\label{tab:table7_3}
\centering

\begin{tabular}{ l | l  l | l  l }
\hline
\multirow{2}{*}{Data} & \multicolumn{2}{c|}{CV} & LS-C & LS-O \\
\cline{2-5}
 & CER & WER & WER & WER \\
\hline
50r & 18.39 & 41.41 & 19.07 & 30.68 \\
100r & 17.10 & 37.87 & 15.29 & 26.81 \\ \hline
50r\_50tts & 20.20 & 40.50 & 16.16 & 27.80 \\
50r\_100tts & 20.00 & 40.28 & 16.15 & 27.82 \\
50r\_200tts & \textbf{19.28} & \textbf{39.85} & \textbf{15.41} & \textbf{27.40} \\
50r\_400tts & 19.97 & 41.32 & 15.91 & 28.59 \\
\hline
\end{tabular}
\end{table}

\section{Conclusion}
\label{conclusion}
This work explored different data augmentation methods to increase the diversity of TTS and VC-generated data that is combined with real data for ASR training. By combining the data augmentation methods that individually improve ASR performance, we were able to reduce the CER and WER of the ASR model trained with real and TTS-augmented data.
In particular, our experiments showed that adding some individual attributes to the training data via TTS synthesis increases ASR performance. In particular, increasing the phonetic diversity of the dataset reduces the CER by 12\% and WER by 14\% relative compared to a dataset containing only real data. In addition, adding new speakers reduces the CER by 5\% and WER by 3\%  relative compared to a dataset of real and TTS-generated data that contains only the speakers in the real dataset. Increasing the phoneme duration diversity relatively reduces the WER by 1\% relative compared to a baseline dataset of real and TTS-generated data.
VC conversion was not as beneficial as TTS for ASR data augmentation. Also, increasing the pitch diversity did not improve ASR performance, probably because pitch information is already adequate in the real data of 2{,}457 speakers. Adding noise and reverb to the TTS-generated dataset combined with the real data for ASR training reduces both the CER and WER by 7\% relative compared to a baseline that does not add noise and reverb.
Finally, combining all the attributes that help improve ASR performance seems to be beneficial and complementary. 
Therefore, this work further highlights the need for diversity of training utterances for better ASR performance, and this diversity can be provided using TTS-generated utterances. We also observe that self-supervised models provide better and robust representations for ASR data augmentation using synthetic data. Controlling diversity is also not limitless as our work shows, and too much diversity can saturate the performance of the ASR system.

\bibliographystyle{IEEEtran}
\bibliography{biblio}



\end{document}